\documentclass[11pt]{article}
\usepackage[margin=1in]{geometry}
\usepackage{setspace}
\usepackage{dsfont, amssymb,amsmath,amscd,latexsym, amsthm, amsxtra,amsfonts, extarrows}
\usepackage{amssymb}
\usepackage{amsmath}
\numberwithin{equation}{section}
\numberwithin{figure}{section}
\usepackage{indentfirst}
\usepackage{graphicx}
\usepackage[colorlinks=true, allcolors=blue]{hyperref}
\usepackage[utf8]{inputenc} 
\usepackage[T1]{fontenc}    
\usepackage{url}            
\usepackage{booktabs}       
\usepackage{amsfonts}       
\usepackage{nicefrac}       
\usepackage{microtype}      
\usepackage{fancyhdr}       
\usepackage{natbib}
\graphicspath{{media/}}     

\newtheorem{Definition}{Definition}

\newtheorem{Remark}{Remark}

\newtheorem{Proposition}{Proposition}
\newtheorem{Assumption}{Assumption}
\usepackage{appendix}
\usepackage{subcaption}

\usepackage{afterpage}
\usepackage{rotating}

\begin{document}

\title{\textbf{Robust Insurance Pricing and Liquidity Management}} 

\author{Shunzhi Pang\thanks{\ School of Economics and Management, Tsinghua University; Faculty of Economics and Business, Katholieke Universiteit Leuven; Email: psz22@mails.tsinghua.edu.cn}}

\date{\today}
\maketitle

\begin{abstract}
\noindent With the rise of emerging risks, model uncertainty poses a fundamental challenge in the insurance industry, making robust pricing a first-order question. This paper investigates how insurers’ robustness preferences shape competitive equilibrium in a dynamic insurance market. Insurers optimize their underwriting and liquidity management strategies to maximize shareholder value, leading to equilibrium outcomes that can be analytically derived and numerically solved. Compared to a benchmark without model uncertainty, robust insurance pricing results in significantly higher premiums and equity valuations. Notably, our model yields three novel insights: (1) The minimum, maximum, and admissible range of aggregate capacity all expand, indicating that insurers’ liquidity management becomes more conservative. (2) The expected length of the underwriting cycle increases substantially, far exceeding the range commonly reported in earlier empirical studies. (3) While the capacity process remains ergodic in the long run, the stationary density becomes more concentrated in low-capacity states, implying that liquidity-constrained insurers require longer to recover. Together, these findings provide a potential explanation for recent skepticism regarding the empirical evidence of underwriting cycles, suggesting that such cycles may indeed exist but are considerably longer than previously assumed. 
\\
\vspace{0in}
\bigskip
\end{abstract}

\newpage

\section{Introduction} 

Model uncertainty, also known as ambiguity, is a salient feature of the insurance industry \citep{knight1921risk, ellsberg1961risk, hogarth1989risk}. While risk pertains to stochastic events with known probability distributions, ambiguity arises when probability distributions are unknown, often referred to as ``unknown unknowns.'' In principle, insurers rely on statistical modeling and actuarial fairness principles to price insurance products. However, for catastrophic risks (e.g., earthquakes, hurricanes, and floods) or emerging risks (e.g., climate and cyber risks), the low frequency of events and the proprietary nature of data pose significant challenges to the accumulation of sufficient data for statistical modeling. Another defining characteristic of risks subject to model uncertainty is their propensity to generate correlated losses, thereby undermining the effectiveness of diversification through the law of large numbers \citep{hogarth1992pricing}. For instance, hurricanes can cause widespread property destruction within a region, while a cyber-attack on cloud computing services may simultaneously disrupt multiple firms relying on the same infrastructure. 

In practice, underwriting risks subject to model uncertainty often leads to substantial financial losses for insurers, prompting them to withdraw from offering coverage in subsequent periods \citep{kunreuther1993insurer}. A recent notable example is the COVID-19 insurance policies sold by major Chinese property and casualty insurers in 2021-2022. These policies, which provided one-year coverage and included a payout of 20,000 RMB upon a confirmed COVID-19 diagnosis, were priced at merely 69 RMB. However, the unexpected surge in infections at the end of 2022 triggered an overwhelming volume of claims, resulting in severe financial distress for insurers, with some even refusing to honor payouts. This underscores the critical challenge insurers encounter in practice. To mitigate market failures, a rigorous theoretical framework is required to guide insurers in developing robust pricing and underwriting strategies.  

Building on recent advances that integrate decision-making under ambiguity aversion with robust control theory \citep{hansen2001robust, hansen2008robustness}, this study examines how insurers’ robustness preferences shape the competitive equilibrium of the insurance market, particularly in contrast to the benchmark case without model uncertainty. We adopt the continuous-time framework developed by \citet{henriet2016dynamics} to model a competitive insurance market subject to financial frictions. Unlike the classical actuarial approach, where insurance premiums are exogenously determined by a safety loading factor or a predefined function of losses, this framework endogenizes the loading through the market-clearing condition, leading to dynamic fluctuations in insurance prices. This feature better captures the underwriting cycles observed in practice \citep{cummins1987international, lamm1997international, chen1999underwriting, harrington2003property}. 

In the benchmark setting, each insurer decides its underwriting, dividend payout, and refinancing strategies to maximize expected shareholder value under the physical probability measure. However, when accounting for model uncertainty, insurers acknowledge the existence of alternative models and incorporate distorted probability measures into their decision-making process. This transforms the optimization problem into a robust control framework, where the underwriting scale, liquidity management policy, and the worst-case measure are jointly determined. The entropy cost associated with model uncertainty follows the specification in \citet{maenhout2020generalized} and \citet{ling2021robust}, providing a structured trade-off between maximizing firm value and ensuring robustness. 

As insurers in our setting are assumed to be homogeneous differing only in initial capital endowments, their optimal underwriting and liquidity management strategies are likewise homogeneous in equilibrium. The industry dynamics are thus governed by the aggregate liquid reserves, which follow a barrier-type structure. When reserves reach a sufficiently high level, insurers distribute dividends to shareholders until capacity falls back below the payout boundary. Conversely, when reserves drop below a specified threshold, insurers raise external capital to restore reserves at least up to that boundary. Between these two boundaries lies the internal financing region, where both the insurance price and the market-to-book ratio are endogenously determined as functions of aggregate capacity. These quantities are jointly characterized by a system of ordinary differential equations derived from the Hamilton-Jacobi-Bellman-Isaacs (HJBI) framework.

The equilibrium outcomes are obtained numerically under reasonable parameter specifications. For any given capacity level, robust pricing leads to substantially higher premiums than the benchmark without model uncertainty, which in turn leads to a notable reduction in underwriting volume. This result is consistent with early survey-based empirical evidence \citep{kunreuther1995ambiguity, cabantous2007ambiguity, cabantous2011imprecise} and corroborates more recent theoretical predictions by \citet{dietz2019ambiguity} and \citet{dietz2021pricing}. Quantitatively, we find that the premium per unit of risk rises by approximately 4.2\%-5.6\%. Furthermore, a higher ambiguity aversion coefficient (or equivalently, a lower level of trust in the physical model), and a higher external financing cost (i.e., the cost of capital in the insurance sector) both contribute to further elevating the equilibrium price. Relative to the benchmark, the market-to-book ratio of insurers in the robust equilibrium is also significantly higher at the same capacity level. This indicates that insurers demand a higher valuation of equity as compensation for bearing the risk of model misspecification. Put differently, robustness concerns effectively raise the shadow cost of capital.

Compared with the discrete-time framework, the main advantage of our continuous-time setting is that it allows us to analyze the impact of model uncertainty on the insurance sector from an evolutionary perspective. We highlight several novel predictions that emerge from our results. First, in the presence of model uncertainty, the minimum, maximum, and admissible range of aggregate capacity all expand. The refinancing boundary is no longer zero but strictly positive, indicating that insurers must maintain higher precautionary reserves. Also, the higher payout boundary and wider capacity range imply that dividend distributions are delayed and that the market can sustain a greater accumulation of liquid reserves. Together, these features highlight that insurers’ liquidity management becomes more conservative and cautious when model uncertainty is considered. 

Second, while the insurance capacity and pricing dynamics in the robust equilibrium still exhibit cyclical patterns, the expected length of the underwriting cycle increases markedly. Specifically, both the soft market phase (i.e., capacity rising from the lower to the upper boundary, with falling prices and rising volumes) and the hard market phase (i.e., capacity declining from the upper to the lower boundary, with rising prices and shrinking volumes) are prolonged. In our benchmark versus robust equilibrium comparison, the expected soft market duration increases from 4.78 to 14.05 years, while the hard market duration rises from 4.84 to 11.92 years, leading the total cycle length to expand from 9.62 to 25.97 years. Both stronger ambiguity aversion and greater financial frictions would further extend the cycle length. Such long durations far exceed the commonly reported 6-10 year range in earlier empirical studies \citep{harrington2013insurance}, and may help rationalize the recent findings by \citet{boyer2012underwriting} and \citet{boyer2015underwriting} that statistical evidence of underwriting cycles in historical data is weak and fragile. If cycles are indeed much longer than previously assumed, then short data samples would fail to contain enough realizations to deliver robust statistical inference. Moreover, introducing model uncertainty renders the cycle more asymmetric. Under plausible parameterizations, the soft market persists substantially longer than the hard market, consistent with the empirical stylized fact emphasized by \citet{henriet2016dynamics}. 

Third, we find that the capacity process is ergodic, converging to a stationary distribution in the long run. Relative to the benchmark case without model uncertainty, the stationary density shifts upward at low capacity levels and downward at high capacity levels. As ambiguity aversion strengthens, the distribution becomes increasingly concentrated near the lower boundary, indicating that insurers are more likely to remain in low-capacity states under stronger robustness concerns. Such result also offers a structural explanation for the empirical difficulty of forecasting underwriting performance documented by \citet{boyer2012underwriting}. If the stationary distribution itself is skewed toward low-capacity states and recovery from adverse shocks is slow, then short-term predictability of underwriting ratios, as tested in their out-of-sample forecasting exercises, will necessarily be limited. In other words, prolonged persistence in depressed states implies that cycles exist in theory, but manifest only weakly in finite historical samples. 

Our paper contributes to several strands of literature. First, it builds on the expanding body of research on model uncertainty in decision-making, particularly in the context of insurance and actuarial science. While much of the recent literature has focused on the implications of ambiguity for insurers’ investment and reinsurance decisions \citep{yi2013robust, zeng2016robust, guan2019robust, li2019optimal, cheng2024robust}, relatively few studies have examined its impact on insurance pricing and market equilibrium, despite its first-order importance. Several studies have established theoretical frameworks for robust insurance pricing. For instance, \citet{zhao2011ambiguity}, \citet{pichler2014insurance}, and \citet{dietz2019ambiguity} propose different criteria for pricing under ambiguity; however, they do not incorporate demand-side dynamics or market-clearing conditions. \citet{anwar2012competitive} analyze the effects of ambiguity on buyer and seller behavior in competitive markets, yet their model remains static and does not capture the dynamic evolution of insurance prices and underwriting cycles. Our study extends this literature by embedding robust pricing into a dynamic market equilibrium framework, offering new insights into its practical applications. 

Second, we contribute to the growing literature on the financial economics of insurance, which emphasizes the role of insurers as financial intermediaries in capital markets and calls for a supply-side theory of insurance \citep{koijen2023financial}. While existing studies have demonstrated that financial frictions, product market power, and statutory reserve regulations impose shadow costs on insurance prices \citep{koijen2015cost, koijen2022fragility, ge2022financial}, we identify an additional factor: insurers' ambiguity aversion stemming from limited knowledge of risk. 

Third, we contribute to the well-established literature on explanations for insurance underwriting cycles, including capacity constraints \citep{gron1994capacity, gron1994evidence}, reporting and regulatory lags \citep{cummins1987international, eckles2016theory}, and information asymmetry \citep{dicks2022asymmetric}. The cycle generated in our model belongs to the first category, namely capacity-driven fluctuations, but its duration is much longer than previous theoretical and empirical estimations. This provides a potential explanation for the ongoing debate on the actual existence of underwriting cycles \citep{boyer2012underwriting, boyer2015underwriting}: cycles may indeed exist, but are stretched significantly by robustness concerns and financial frictions. As a result, within a conventional 30-40 year data window, fewer than two complete cycles may be observed, making statistical identification extremely difficult. Hence, rather than relying on short-term cycle predictions, we advocate that insurers adopt robust pricing strategies to mitigate the risk of underwriting losses.

Last, our work connects to the classical literature on liquidity management in financial intermediation. While insurers can raise debt financing at relatively low costs through policyholder reserves, equity remains expensive, similar to other financial intermediaries \citep{brunnermeier2009market, he2013intermediary}. Accumulated equity capital plays a critical role in shaping insurers' behavior \citep{winter1994dynamics, henriet2016dynamics, luciano2022fluctuations}. However, unlike banks or funds, insurers are exposed not only to financial risks but also to physical risks. The occurrence of an extreme physical loss can substantially reduce insurers’ liquid assets and may even force them to fire-sell long-term investments, thereby heightening the risk of systemic spillovers to financial markets. In the post-pandemic era, we highlight the importance for financial intermediaries to incorporate model uncertainty into their risk and liquidity management frameworks. 

The remainder of the paper is organized as follows. Section \ref{Section Model} introduces model uncertainty and defines the market equilibrium. Section \ref{Section Equilibrium} presents insurers’ optimization problem and derives the theoretical results. Section \ref{Section Numerical} provides a numerical analysis of equilibrium outcomes and compares the cases with and without model uncertainty. Section \ref{Section Long-Run Behavior} examines insurers’ long-run behavior, focusing on the duration of underwriting cycles and the ergodic property of capacity dynamics. Finally, Section \ref{Section Conclusion} concludes the paper.

\section{General Model and Problem Formulation} \label{Section Model}

In this section, we build upon the theoretical framework of \citet{henriet2016dynamics} and extend it to incorporate insurers' concerns about model uncertainty, following the formulation of \citet{hansen2001robust} and \citet{anderson2003quartet}. As we shall demonstrate, accounting for model uncertainty introduces additional complexity into the equilibrium analysis, ultimately shaping both insurers' underwriting strategies and the equilibrium pricing of insurance. 

Let $(\Omega,\mathcal{F},\{\mathcal{F}_t\}_{t\geq0},\mathbb{P})$ be a complete probability space satisfying the usual conditions. All stochastic processes governing the insurance market in the subsequent discussion are assumed to be well-defined on this space. 

\subsection{Insurance Market} 

We consider a competitive insurance market consisting of a continuum of insurers, each providing coverage to individuals exposed to perfectly correlated risks. Idiosyncratic risks at the individual level are disregarded, as they can be fully diversified in a large population and thus have no impact on aggregate outcomes.\footnote{This setting aligns with the feature of risks subject to model uncertainty, their tendency to generate correlated losses, as discussed in the introduction. However, we acknowledge that studying a market with partially correlated risks is a meaningful extension. } The cumulative loss process for a representative insurer, denoted by $L \triangleq \{ L_t: t \geq 0 \}$, evolves as the following dynamics:
\begin{equation}
    \mathrm{d}L_t = l \mathrm{d}t + \eta \mathrm{d}B_t, \label{Benchmark Loss Process}
\end{equation}
where $l > 0$ represents the expected instantaneous loss rate, $\eta > 0$ measures the insurer’s exposure to systematic loss risk, and $B \triangleq \{ B_t: t \geq 0 \}$ is a one-dimensional standard Brownian motion. 

Unlike traditional insurance markets, where insurers rely on extensive underwriting experience and historical data, the risks considered here belong to the category of “unknown unknowns”, where insurers possess limited data to form reliable probabilistic beliefs. Such risks are prevalent in practice, including catastrophic events \citep{jaffee1997catastrophe}, large-scale pandemics \citep{fan2017pandemic}, and emerging cyber threats \citep{eling2016we}, where historical data are insufficient for precise risk assessment. Given their strong interdependencies, it is natural to model these them as systematic risks subject to model uncertainty. 

Insurance contracts are assumed to be short-term. That is, for a continuously evolving risk process, each policy expires within an infinitesimal time interval, allowing both the insuree and the insurer to dynamically adjust their risk exposure. Such setting has been widely adopted in the literature on dynamic insurance markets \citep[e.g.,][]{henriet2016dynamics, luciano2022fluctuations}, and is particularly suitable for analyzing property and casualty (P\&C) insurance markets, where contracts are typically of short maturity and insurers do not accumulate long-term liabilities. Assume the insurance premium per unit of time is given by:  
\begin{equation}
    \pi_t \mathrm{d}t = (l + \eta p_t) \mathrm{d}t, \label{Premium}
\end{equation}
which consists of two components: the actuarially fair premium $l$, and a loading premium, determined by the risk exposure $\eta$ and a loading factor $p_t$. The loading factor $p_t$ is endogenously determined through market-clearing between supply and demand. For simplicity, we refer to $p_t$ as the price of insurance in what follows. 

In the literature, there are different approaches to characterize insurance market demand. Earlier classical studies examine optimal insurance and risk sharing within the framework of expected utility maximization, either in an intertemporal or life-cycle context \citep[e.g.,][]{yaari1965uncertain, rothschild1976equilibrium, campbell1980demand, cummins2004demand}. More recent contributions incorporate behavioral factors, such as rank-dependent utility, prospect theory, and ambiguity aversion \citep[e.g.,][]{bernard2015optimal, schmidt2016insurance, peter2020you}.

In \citet{henriet2016dynamics}, the market demand is exogenously specified by a continuously differentiable and decreasing function $D(p)$, without imposing any more restrictions on its functional form. While such a flexible setting is also feasible within our framework, we instead adopt the approach of \citet{luciano2022fluctuations} and endogenize the demand function, which has the advantage of yielding a complete general equilibrium model. Specifically, assume that a representative insuree faces the unit loss process \eqref{Benchmark Loss Process}, and transfers a fraction $d_t$ of the risk to the insurer by purchasing insurance at premium $\pi_t$, while retaining the remaining fraction $1 - d_t$. Under mean-variance preferences over an interval $\mathrm{d}t$, the optimal coverage is given by $d^{\ast}_t = 1 - \frac{1}{\alpha \eta} p_t$, where $\alpha > 0$ denotes the degree of risk aversion.\footnote{The detailed derivation can be found in Appendix A of \citet{luciano2022fluctuations}. To simplify the analysis, we do not impose the restriction $d_t \in [0,1]$ at this stage.} Hence, the demand function can be expressed as:
\begin{equation}
    D(p) = 1 - \tfrac{1}{\alpha \eta} p. 
\end{equation}
When $p = 0$, this corresponds to full insurance demand. In what follows, we shift our focus to the characterization of the supply side.

Since insurance contracts are assumed to have an infinitesimal duration, insurers bear no long-term liabilities and are therefore not required to hold reserves against them.\footnote{This assumption departs from actual practice, as even P\&C insurers are typically required to hold additional reserves to meet solvency regulations. Here, we adopt this simplification to enhance analytical tractability.} On each insurer's balance sheet, the cash or liquid reserves on the asset side, denoted by $m \triangleq \Big\{m_t: t \geq 0 \Big\}$, should equal the equity $e \triangleq \Big\{e_t: t \geq 0 \Big\}$. Thus, $m_t$ simultaneously represents the book value of equity for an insurer and can also be interpreted as the net wealth of the insurer. 

Suppose an insurer's underwriting scale process is given by $x \triangleq \Big\{x_t \geq 0: t \geq 0 \Big\}$. Over an infinitesimal period $(t, t + \mathrm{d}t)$, the insurer collects premium income of $x_t \pi_t \mathrm{d}t$ and incurs claim payouts of $x_t \mathrm{d}L_t$. The difference $x_t \big( \pi_t \mathrm{d}t - \mathrm{d}L_t \big)$, represents the underwriting profit, which can either be retained as cash reserves or distributed as dividends. Let $\delta \triangleq \Big\{\delta_t: t \geq 0 \Big\}$ denote the insurer's cumulative dividend process. The insurer optimally determines the dividend payout $\mathrm{d}\delta_t \geq 0$ to distribute to shareholders. Additionally, in the event of financial distress, the insurer can raise external capital to replenish cash reserves. Let $i \triangleq \Big\{i_t: t \geq 0 \Big\}$ represent the cumulative recapitalization process, where $\mathrm{d}i_t \geq 0$. Then, the insurer's net wealth process evolves according to: 
\begin{equation}
    \mathrm{d} m_t = x_t \big( \pi_t \mathrm{d}t - \mathrm{d}L_t \big) + \mathrm{d} i_t - \mathrm{d}\delta_t = x_t \eta p_t \mathrm{d}t - x_t \eta \mathrm{d}B_t + \mathrm{d}i_t - \mathrm{d}\delta_t. \label{Benchmark Individual Wealth Process}
\end{equation}

Due to asymmetric information and managerial incentive problems, the corporate finance literature widely acknowledges that external financing is costly for firms.\footnote{There is extensive theoretical and empirical evidence on this topic, see \citet{jensen1976theory}, \citet{ leland1977informational}, \citet{myers1984corporate}. } The same applies to financial intermediations such as insurers, where regulatory constraints and capital requirements further elevate financing costs \citep{cummins1994capital, cummins2005estimating}. Let $\gamma > 0$ denote the per-unit cost of raising external capital, which would be incorporated into the insurer’s financial evaluation. Finally, we assume that all insurers start with a positive liquid reserve level, i.e., $m^j_0 > 0$ for all $j \in \mathcal{J}$.\footnote{This assumption ensures that, in equilibrium, the liquid reserve level of all insurers will maintain the same sign at any point in time, as will be demonstrated later. } 

\subsection{Optimization Objective without Model Uncertainty}

Let $\mathcal{J} \triangleq [0, 1]$ denote the set of insurers, each indexed by $j \in \mathcal{J}$. Given the liquid reserve process of each individual insurer $m^j$, the aggregate liquid reserves (capacity) of the entire insurance sector are defined as $M \triangleq \Big\{ M_t = \int_{\mathcal{J}} m^j_t \mathrm{d}j: t \geq 0 \Big\}$, which follow the process: 
\begin{align}
    \mathrm{d}M_t & = X_t \eta p_t \mathrm{d}t - X_t \eta \mathrm{d}B_t + \mathrm{d}I_t - \mathrm{d}\Delta_t \notag \\
    & = D(p_t) \eta p_t \mathrm{d}t - D(p_t) \eta \mathrm{d}B_t + \mathrm{d}I_t - \mathrm{d}\Delta_t,  \label{Benchmark Aggregate Wealth Process}
\end{align}
where $X_t = \int_{\mathcal{J}} x^j_t \mathrm{d}j$, $I_t = \int_{\mathcal{J}} i^j_t \mathrm{d}j$ and $\Delta_t = \int_{\mathcal{J}} \delta^j_t \mathrm{d}j$, represent the aggregate underwriting scale, cumulative recapitalization and dividend payouts, respectively. In equilibrium, the aggregate supply must equal market demand, leading to the market-clearing condition $X_t = D(p_t)$. 

Without considering model uncertainty, \citet{henriet2016dynamics} solve the Markovian stationary equilibrium, where the price of insurance is a deterministic function of the aggregate level of reserves, i.e., $p_t = p(M_t)$. Each individual insurer takes the dynamics \eqref{Benchmark Aggregate Wealth Process} and the equilibrium price function $p(M)$ as given and optimally chooses its underwriting scale $x \geq 0$, recapitalization policy $\mathrm{d}i \geq 0$, and dividend policy $\mathrm{d}\delta \geq 0$ to maximize shareholder value: 
\begin{equation}
    v(m, M) = \max_{x \geq 0, \mathrm{d}\delta \geq 0, \mathrm{d}i \geq 0} \mathbb{E} \left\{ \int_0^{\infty} e^{-r t} \Big[ \mathrm{d}\delta_t - (1 + \gamma) \mathrm{d} i_t \Big] \right\}, \label{Benchmark Optimization Objective}
\end{equation}
where $r > 0$ represents the discount factor, such as the risk-free rate.  

\begin{Definition}
    In the absence of model uncertainty concerns, a stationary Markovian competitive equilibrium consists of an aggregate liquid reserve process $M$, a market price of insurance $p(M)$, insurance supply functions $x^j(M), j \in \mathcal{J}$, that are compatible with each insurer's optimization objective \eqref{Benchmark Optimization Objective} and the market-clearing condition $\int_{\mathcal{J}} x^j(M) \mathrm{d}j = D(p(M))$. 
\end{Definition}

\subsection{Optimization Objective with Model Uncertainty} 

Now, we consider the case where insurers are concerned about model uncertainty in the loss process, which is driven by the Brownian motion $B$ under the physical measure $\mathbb{P}$. The insurer acknowledges the possibility of alternative models and considers distorted probability measures that are mutually absolutely continuous with respect to $\mathbb{P}$ over any finite time interval. To formalize this, let $h$ be a density generator associated with the loss process, and denote the set of all such generators by $\mathcal{H}$. Given $h \in \mathcal{H}$, we define process $\xi^h \triangleq \Big\{\xi^h_t: t \geq 0 \Big\}$ as follows: 
\begin{equation}
    \xi_t^h = \exp \left( - \int_0^t h_s \mathrm{d}B_s - \frac{1}{2} \int_0^t h_s^2 \mathrm{d}s \right). \notag
\end{equation}
It is assumed that $\int_0^t h_s^2 \mathrm{d}s < \infty$ for any $t > 0$, ensuring that $\xi^h$ is a martingale process. 

By Girsanov’s Theorem, there exists a subjective probability measure $\mathbb{Q}^h$ such that $\frac{\mathrm{d}\mathbb{Q}^h}{\mathrm{d} \mathbb{P}} \mid_{\mathcal{F}_t} = \xi_t^h$. Under this new measure, the process $B^h \triangleq \Big\{B^h_t: t \geq 0 \Big\}$ defined by: 
\begin{equation}
    \mathrm{d}B_t^h = \mathrm{d}B_t + h_t\mathrm{d}t, \notag
\end{equation}
is a standard Brownian motion. Correspondingly, the loss process \eqref{Benchmark Loss Process} can be rewritten as: 
\begin{equation}
    \mathrm{d}L_t = l \mathrm{d}t + \eta \left(\mathrm{d}B^h_t - h_t\mathrm{d}t \right) =  (l - \eta h_t) \mathrm{d}t + \eta \mathrm{d}B_t^h. \notag
\end{equation}
Similarly, the individual liquid reserve process \eqref{Benchmark Individual Wealth Process} evolves as: 
\begin{equation}
    \mathrm{d}m_t = x_t \left( \pi_t \mathrm{d}t - \mathrm{d}L_t \right) + \mathrm{d}i_t - \mathrm{d}\delta_t = x_t  \eta \left[ \left( p_t + h_t \right) \mathrm{d}t - \mathrm{d}B_t^h \right] + \mathrm{d}i_t - \mathrm{d}\delta_t. \notag
\end{equation} 
Here, the actuarially fair premium used for pricing remains $l$, as it serves as the benchmark accepted by policyholders who do not exhibit preferences for robustness. Finally, from the insurer’s perspective, the dynamics of aggregate capacity for the entire insurance sector are given by: 
\begin{equation}
    \mathrm{d} M_t = X_t \eta (p_t + h_t) \mathrm{d}t - X_t \eta \mathrm{d}B_t^h  + \mathrm{d}I_t - \mathrm{d}\Delta_t. \notag
\end{equation}
Notice that the density generators chosen by individual insurers do not necessarily have to be consistent across different insurers. But as we focus on an equilibrium where insurers are homogeneous in behavior, we pre-assume and later verify that the optimally chosen $h^j$ should be the same value. 

With model uncertainty, each insurer's optimization objective deviates from \eqref{Benchmark Optimization Objective} and is evaluated under the subjective measure $\mathbb{Q}^h$, incorporating a penalty term to account for robustness against uncertainty aversion. We adopt the variational preferences framework of \citet{anderson2003quartet} and \citet{maccheroni2006ambiguity, maccheroni2006dynamic}, where an insurer's optimization objective is given by: 
\begin{equation}
    \inf_{\mathbb{Q}^h \in \mathcal{Q}} \mathbb{E}^{\mathbb{Q}^h} \left\{ \int_0^{\infty} e^{-rt} \Big[ \mathrm{d}\delta_t - (1 + \gamma) \mathrm{d} i_t \Big] \right\} + \mathcal{K}(\mathbb{Q}^h), \label{Robust Optimization Objective}
\end{equation}
where $\mathcal{Q}$ denotes the set of all admissible probability measures $\mathbb{Q}^h$, and $\mathcal{K}(\mathbb{Q}^h)$ is a penalty term that quantifies the entropy cost of model uncertainty, specified as: 
\begin{equation}
    \mathcal{K}(\mathbb{Q}^h) = \frac{1}{2} \mathbb{E}^{\mathbb{Q}^h} \left[ \int_0^\infty e^{-rt} \Theta_t h_t^2 \mathrm{d}t \right]. \notag
\end{equation}
following \citet{maenhout2020generalized} and \citet{ling2021robust}.\footnote{This formulation is equivalent to $\mathcal{K}(\mathbb{Q}^h) = \mathbb{E}^{\mathbb{P}}\left[\int_0^\infty e^{-rt} \Theta_t \mathrm{d}\phi\left(\frac{\xi^h_t}{\xi^h_0}\right) \right]$, where $\phi(x) \triangleq x \ln x$; see \citet{maenhout2020generalized} for technical details. } It represents a discounted, state‐weighted relative entropy between $Q^{h}$ and the reference model $\mathbb{P}$, with $\Theta_t$ determining the cost of distortion: larger $\Theta_t$ makes deviations more expensive (less ambiguity), while smaller $\Theta_t$ permits stronger worst‐case tilting. The value function is denoted as $v(m, M)$, jointly determined by two state variables, namely the individual liquid reserves $m$ and the aggregate capacity $M$. We aim to eliminate one of them to simplify the solution to the control problem. 

For tractability, we assume that $\Theta_t(m_t) = \theta m_t$, where $\theta > 0$ measures the degree of concern for robustness among all insurers.\footnote{Equivalently, $\frac{1}{\theta}$ represents the degree of ambiguity aversion \citep{maccheroni2006ambiguity, maccheroni2006dynamic}. A larger $\theta$ implies lower ambiguity aversion among insurers. } This assumption has two implications. First, the entropy cost is proportional to the insurer’s liquid reserves, which is economically intuitive since larger insurers face higher uncertainty-related costs. Second, such specification ensures that the optimization objective \eqref{Robust Optimization Objective} remains homogeneous in $(m, x, \mathrm{d}\delta, \mathrm{d}i)$, implying the value function should be also linear in individual insurers' net wealth. 

Based on this, we assume that the value function takes the form $v(m, M) = m u(M)$, where $u(M)$ ($m \neq 0$) represents the market-to-book value of each insurer, which is identical across the entire insurance sector. It is conceptually equivalent to the insurance analogue of Tobin’s q ratio, as discussed in \citet{winter1994dynamics} and \citet{henriet2016dynamics}. 

\begin{Definition}
    In the presence of model uncertainty concerns, a stationary Markovian competitive equilibrium consists of an aggregate liquid reserve process $M$, a market price of insurance $p(M)$, insurance supply functions $x^j(M), j \in \mathcal{J}$, that are compatible with each insurer's optimization objective \eqref{Robust Optimization Objective} and the market-clearing condition $\int_{\mathcal{J}} x^j(M) \mathrm{d}j = D(p(M))$. 
\end{Definition}

\section{Equilibrium Solution} \label{Section Equilibrium}

In this section, we derive the insurer’s optimal underwriting and liquidity management strategies by solving the robust control problem. We establish the conditions that must be satisfied by the insurance price and market-to-book value. We then analyze the equilibrium properties, emphasizing its key differences from the benchmark case without model uncertainty. 

\subsection{Benchmark Case}

For a benchmark comparison, we first recall the market equilibrium without concerns of model uncertainty, as established by \citet{henriet2016dynamics}. We omit the detailed derivation, and interested readers may refer to their work for further technical details. 

\begin{Proposition}[\textbf{Equilibrium without Model Uncertainty}] \label{Proposition Benchmark Equilibrium}
    In the absence of model uncertainty, suppose that the following system of equations admits a solution for $p^{\ast}(M)$ and $u^{\ast}(M)$ on $[0, \overline{M}]$: 
    \begin{equation}
        \left\{
        \begin{aligned}
            & p(M) = -\frac{u^{\prime}(M)}{u(M)}D(p(M))\eta, \\
            & 2r = \left[\frac{u^{\prime \prime}(M)}{u(M)} - 2{\left(\frac{u^{\prime}(M)}{u(M)}\right)}^2\right] D(p(M))^2 \eta^2, 
        \end{aligned}
        \right. \label{Solution to Benchmark Equilibrium} 
    \end{equation}
    subject to the demand function: $D(p(M)) = 1 - \frac{1}{\alpha \eta} p(M)$, and boundary conditions: $u(0) = 1 + \gamma$, $u(\overline{M}) = 1$, and $u^{\prime}(\overline{M}) = 0$. Then, there exists a stationary Markovian equilibrium such that: 

    (1) For $M \in (0, \overline{M})$, the market price of insurance is $p^{\ast}(M)$, and each insurer's market-to-book value function is $u^{\ast}(M)$. 

    (2) For $M \geq \overline{M}$, insurers distribute dividends to shareholders until the aggregate liquid reserves fall below $\overline{M}$. For $M \leq 0$, insurers raise external capital to restore positive reserves. 
\end{Proposition} 

\begin{Proposition} \label{Proposition Benchmark Properties}
    The equilibrium insurance price $p^{\ast}(M)$ is strictly decreasing in the aggregate capacity $M$. The upper bound is given by $p^{\ast}(0)$, while the lower bound satisfies $p^{\ast}(\overline{M}) = 0$. Consequently, the equilibrium price remains non-negative for all $M \in [0, \overline{M}]$, ensuring that insurers’ expected profits from underwriting remain non-negative. 
\end{Proposition} 

As shown, the aggregate capacity of the insurance industry $M$ serves as the unique state variable determining both the equilibrium insurance price and the market-to-book ratio. In a dynamic setting, as $M$ fluctuates, insurance prices exhibit corresponding volatility, leading to underwriting cycles. The insurer’s decision-making problem results in a three-region structure for the state $M$: an external financing region ($M \leq \underline{M}$), an internal financing region ($\underline{M} < M < \overline{M}$), and a payout region ($M \geq \overline{M}$). Such barrier-type solution has been widely applied in liquidity management problems; see \citet{rochet2011liquidity} and \citet{bolton2011unified}. Next, we extend this solution structure to the case with model uncertainty. 

\subsection{Internal Financing Region} 

In the presence of model uncertainty concerns, an insurer's value function $v(m, M)$ should satisfy the following Hamilton-Jacobi-Bellman-Isaacs (HJBI) equation: 
\begin{align}
    r v(m ,M) & = \sup_{x \geq 0, \mathrm{d}\delta \geq 0, \mathrm{d}i \geq 0} \inf_{h \in \mathcal{H}} \bigg\{ \frac{1}{2} \Theta(m) h^2 + x \eta (p + h) v_m + D(p) \eta (p + h) v_M + \frac{1}{2} x^2 \eta^2 v_{mm} \notag \\
    & + \frac{1}{2} D(p)^2 \eta^2 v_{MM} + x D(p) \eta^2 v_{mM} + \mathrm{d}\delta (1 - v_m) + \mathrm{d}i (v_m - 1 - \gamma)\bigg\}. \label{HJBI Original}
\end{align}
Here, we pre-assume that both $v(\cdot, \cdot)$ and $u(\cdot)$ are smooth functions, with $u(\cdot) > 0$, continuously differentiable up to the second order. Besides, we impose $m \geq 0$, it would be unreasonable to assume that the insurer continues underwriting when its liquid reserves are negative. Actually, when considering a barrier-type solution, the insurer is assumed to recapitalize once $m$ falls below a lower threshold $\underline{M}$, which is typically non-negative.

We begin by analyzing the internal financing region, where the optimal dividend and recapitalization policies are $\mathrm{d}\delta^{\ast} = 0$ and $\mathrm{d}i^{\ast} = 0$. Hence, the focus reduces to the max-min problem for the remaining controls. For $m = 0$, we have $\Theta(m) = 0$, and the minimization over $h$ admits an interior solution if and only if $x^{\ast} = - \frac{v_M}{v_m}D(p)$. Given the assumption that $v(m, M) = m u(M)$, it leads to $x^{\ast}(0, M) = - \left. \frac{m u^{\prime}(M)}{u(M)} \right|_{m = 0} D(p) = 0$. Under this condition, the HJBI equation holds trivially, as both sides are equal to zero. This result hints that regardless of the aggregate capacity $M$, an individual insurer does not engage in underwriting when it has zero liquid reserves. 

For $m > 0$, the minimization over $h$ is attained at: 
\begin{equation}
    h^{\ast} = - \frac{x \eta v_m + D(p) \eta v_M}{\Theta(m)}. \label{Candidate h}
\end{equation}
Substituting it back into the HJBI equation, the first-order condition with respect to $x$ is given by: 
\begin{equation}
    \frac{\partial h^{\ast}}{\partial x} \Big[ \Theta(m)h^{\ast} + x \eta v_m + D(p) \eta v_M \Big] + \eta (p + h^{\ast}) v_m + x \eta^2 v_{mm} + D(p) \eta^2 v_{mM} = 0. \notag
\end{equation}
Under the assumption that $v(m, M) = m u(M)$, the candidate underwriting scale: 
\begin{equation}
    x^{\ast}(m, M) = m \left[ \frac{\theta}{u(M) \eta} \left( p + \frac{u^{\prime}(M)}{u(M)}D(p) \eta \right) - \frac{u^{\prime}(M)}{u(M)}D(p) \right], \label{Optimal x}
\end{equation}
is optimal when it is non-negative. Otherwise, due to the non-negativity constraint, the optimal scale is restricted to $x^{\ast}(m, M) = 0$. 

We analyze the two cases separately. For some $M$ such that $f(M) \triangleq \frac{\theta}{u(M) \eta} \left( p + \frac{u^{\prime}(M)}{u(M)}D(p) \eta \right) - \frac{u^{\prime}(M)}{u(M)}D(p) > 0$, then $x^{\ast}(m, M) > 0$ and the optimal density generator is then given by:
\begin{equation}
    h^{\ast}(M) = - p(M) - \frac{u^{\prime}(M)}{u(M)} D(p(M)) \eta. \label{Optimal h}
\end{equation}
Substituting \eqref{Optimal x} and \eqref{Optimal h} back into the HJBI equation \eqref{HJBI Original} yields: 
\begin{equation}
    2r = \frac{\theta}{u(M)}\left( p(M) + \frac{u^{\prime}(M)}{u(M)} D(p(M)) \eta \right)^2 + \left[\frac{u^{\prime \prime}(M)}{u(M)} - 2{\left(\frac{u^{\prime}(M)}{u(M)}\right)}^2\right] D(p(M))^2 \eta^2. \label{Solution to u}
\end{equation}

However, if $f(M) \leq 0$, $x^{\ast}(m, M) = 0$ for all $m > 0$. Combined with $x^{\ast}(0, M) = 0$ for any $M$, the market-clearing condition implies $D(p^{\ast}(M)) = 0$. Then, the optimal density generator solved by \eqref{Candidate h} is $h^{\ast}(M) = 0$. Substituting these values back into the HJBI equation, we have $r v(m, M) = r m u(M) = 0$, which contradicts the assumption that $m > 0$ and $u(M) > 0$. Thus, this scenario cannot occur. If an equilibrium exists, it must satisfy $f(M) > 0$ for any $M$. 

Based on \eqref{Optimal x} and $x^{\ast}(0, M) = 0$ for any $M$, the market-clearing condition is equivalent to: 
\begin{equation}
    D(p(M)) =  M \left[ \frac{\theta}{u(M) \eta} \left( p(M) + \frac{u^{\prime}(M)}{u(M)} D(p(M)) \eta \right) - \frac{u^{\prime}(M)}{u(M)}D(p(M)) \right], \label{Solution to p}
\end{equation}
where $D(p(M))$ on the left side equals the sum of $x^{\ast}(m, M)$, and the multiplier $M$ on the right side is the sum of $m$ across insurers. Once the specific form of the demand function $D(p)$ is specified, we can express the equilibrium price $p^{\ast}(M)$ as a function of the market-to-book ratio $u(M)$ and its first-order derivative $u^{\prime}(M)$, though their exact values have yet to be determined. Replacing the price function into equation \eqref{Solution to u}, we obtain a second-order ODE for $u(\cdot)$.  

Finally, we note that the optimal underwriting scale derived in \eqref{Optimal x} remains valid for $m = 0$. This implies that each insurer’s underwriting scale is proportional to its liquid reserve level, aligning with the homogeneity assumption that $v(m, M) = m u(M)$. Furthermore, the optimal density generator \eqref{Optimal h} depends solely on the aggregate capacity $M$, implying that all insurers with positive liquid reserves share the same belief regarding the worst-case scenario.  

\subsection{External Financing and Payout Regions} 

Next, we analyze the optimal dividend and recapitalization policies. Based on the HJBI equation \eqref{HJBI Original}, the maximization with respect to $\mathrm{d}\delta$ and $\mathrm{d}i$ is independent of the choice of $h$. A necessary condition for the existence of an equilibrium solution is: 
\begin{equation}
    1 - v_m \leq 0, \quad \text{and} \quad  v_m - 1 - \gamma \leq 0, \notag
\end{equation}
that is $1 \leq u(M) \leq 1 + \gamma$. Moreover, $\mathrm{d}\delta^{\ast} > 0$ only if $u(M) = 1$, while $\mathrm{d}i^{\ast} > 0$ only if $u(M) = 1 + \gamma$. In line with the benchmark case without model uncertainty, as well as many related liquidity management problems, we focus on a barrier-type solution. Specifically, $\overline{M}$ and $\underline{M}$ denote the payout boundary and the external financing boundary, respectively. And to ensure the existence of a solution, we impose the following assumption on the market-to-book ratio function. 
\begin{Assumption}\label{Assumption market-to-book ratio}
    For $M \in [\underline{M}, \overline{M}]$ with $0 \leq \underline{M} < \overline{M}$, it holds that $1 \leq u(M) \leq 1 + \gamma$ and $u^{\prime}(M) \leq 0$. In other words, the market-to-book ratio of the insurance sector is assumed to be a decreasing function of aggregate capacity, bounded between its two boundary values. 
\end{Assumption}

The monotonicity of $u(\cdot)$ is essential for ensuring that the optimal policy takes a barrier-type form. Otherwise, the sets $\{M: u(M) \leq 1\}$ and $\{M: u(M) \geq 1+\gamma\}$ could consist of multiple disconnected intervals, leading to multiple thresholds and more complex strategies.

When liquid reserves are relatively high ($M \geq \overline{M}$), the insurer maximizes shareholder value by distributing excess cash as dividends. At the payout boundary ($M = \overline{M}$), the insurer is indifferent between retaining or distributing one additional dollar, implying $u(\overline{M}) = 1$. This ensures that the marginal value of cash equals the marginal cost to shareholders, preventing excessive accumulation of reserves. Conversely, when liquid reserves fall below a critical threshold ($M \leq \underline{M}$), the insurer optimally raises external funds to restore liquidity. The external financing boundary satisfies $u(\underline{M}) = 1 + \gamma$. It ensures that the marginal value of cash equals the marginal cost of recapitalization, making external financing optimal only when strictly necessary. 

Referring to \citet{henriet2016dynamics} and \citet{luciano2022fluctuations}, we additionally impose a no-arbitrage condition at the boundaries. Let $V(M) = M u(M)$ denote the market value of the entire insurance industry. Then, the absence of arbitrage opportunities requires the following smooth-pasting conditions: 
\begin{align}
    & V^{\prime}(\overline{M}) = u(\overline{M}) + \overline{M} u^{\prime}(\overline{M}) = 1, \notag \\
    & V^{\prime}(\underline{M}) = u(\underline{M}) + \underline{M} u^{\prime}(\underline{M}) = 1 + \gamma. \notag
\end{align}
In words, the marginal change in the aggregate market value induced by a dividend payout or recapitalization must equal the marginal flow of funds withdrawn from or injected into the industry by shareholders. At the upper bound, it derives that $u^{\prime}(\overline{M}) = 0$, as $\overline{M} > 0$. At the lower bound, however, it indicates that $\underline{M} = 0$ or $u^{\prime}(\underline{M}) = 0$. 

The contingent condition $\underline{M} = 0$ cannot generally hold in our setting. Suppose $\underline{M} = 0$ and there exists a solution $\left(p^{\ast}(M), u^{\ast}(M)\right), \, M \in [\underline{M}, \overline{M}]$, to the market equilibrium. Then, the market-clearing condition \eqref{Solution to p} implies $D(p^{\ast}(0)) = 0$ and $p^{\ast}(0) = \alpha \eta$. Since equation \eqref{Solution to u} holds for $M > 0$, it should also hold at $M = 0$, provided that $u(0) \neq 0$ and both $p(\cdot)$ and $u(\cdot)$ are sufficiently smooth and bounded in a neighborhood of $M = 0$. Then, substituting $\underline{M} = 0$ and $D(p^{\ast}(0)) = 0$ into the equation yields: 
\begin{equation}
    2r = \frac{\theta}{1 + \gamma} \alpha^2 \eta^2, \notag
\end{equation} 
which does not generally hold unless specific parameter restrictions are satisfied. Therefore, the alternative condition $u^{\prime}(\underline{M}) = 0, \ \underline{M} > 0$, should be satisfied.
\begin{Remark}
Unlike \citet{henriet2016dynamics} and \citet{luciano2022fluctuations}, where the lower barrier is shown to be zero because the optimally solved $u(\cdot)$ is convex in the aggregate capacity (hence $u^{\prime}(\underline{M}) = 0$ and $u^{\prime}(\overline{M}) = 0$ cannot occur simultaneously), the incorporation of model uncertainty here leads to a different form of solution, and $\underline{M} = 0$ cannot generally hold. In Section \ref{Section Numerical}, we demonstrate through numerical simulations that the market-to-book ratio can satisfy the value-matching and smooth-pasting boundary conditions, as well as the requirements in Assumption \ref{Assumption market-to-book ratio}.
\end{Remark}

\subsection{Market Equilibrium}

Combining the solution to the control problem with the determination of the payout and external financing boundaries, we characterize the market equilibrium in the following proposition. 

\begin{Proposition}[\textbf{Equilibrium with Model Uncertainty}] \label{Proposition Robust Equilibrium}
    In the presence of model uncertainty, suppose that the following system of equations admits a solution for $p^{\ast}(M)$ and $u^{\ast}(M)$ on $[\underline{M}, \overline{M}]$: 
    \begin{equation}
        \left\{
        \begin{aligned}
            & D(p(M)) =  M \left[ \frac{\theta}{u(M) \eta} \left( p(M) + \frac{u^{\prime}(M)}{u(M)} D(p(M)) \eta \right) - \frac{u^{\prime}(M)}{u(M)}D(p(M)) \right] , \\
            & 2r = \frac{\theta}{u(M)}\left( p(M) + \frac{u^{\prime}(M)}{u(M)} D(p(M)) \eta \right)^2 + \left[\frac{u^{\prime \prime}(M)}{u(M)} - 2{\left(\frac{u^{\prime}(M)}{u(M)}\right)}^2\right] D(p(M))^2 \eta^2, 
        \end{aligned}
        \right. \label{Solution to Robust Equilibrium}
    \end{equation}
    subject to demand function: $D(p(M)) = 1 - \frac{1}{\alpha \eta} p(M)$,  boundary conditions: $u(\underline{M}) = 1 + \gamma$, $u(\overline{M}) = 1$, $u^{\prime}(\underline{M}) = u^{\prime}(\overline{M}) = 0$, and the requirements in Assumption \ref{Assumption market-to-book ratio}. Then, there exists a stationary Markovian equilibrium such that: 

    (1) For $M \in (\underline{M}, \overline{M})$, the market price of insurance is $p^{\ast}(M)$, each insurer's market-to-book value function is $u^{\ast}(M)$, and the worst-case density generator is $h^{\ast}(M)$ as given in \eqref{Optimal h}. For the insurer with reserves $m$, its optimal operation amount is $x^{\ast}(m, M)$ as given in \eqref{Optimal x} and the shareholders value is $v^{\ast}(m, M) = m u^{\ast}(M)$. 
    
    (2) For $M \geq \overline{M}$, insurers distribute dividends to shareholders until the aggregate liquid reserves fall below $\overline{M}$. For $M \leq \underline{M}$, insurers raise external capital to restore reserves to the level of $\underline{M}$.
\end{Proposition} 

Combining the first equation in \eqref{Solution to Robust Equilibrium} with the demand function, we have the representation: 
\begin{equation}
    p(M) = \frac{g(M)}{M \frac{\theta}{u(M)\eta} + \frac{1}{\alpha \eta} g(M)}, \quad \text{where} \quad g(M) = 1 - M \frac{u^{\prime}(M)}{u(M)} \left( \frac{\theta}{u(M)} - 1\right). \notag
\end{equation}
Substituting this into the second equation yields a second-order ODE for $u(\cdot)$: 
\begin{equation}
    u^{\prime \prime}(M) = \frac{2 {u^{\prime}}^2(M)}{u(M)} + 2r u(M) \left(\frac{1}{\eta} + \frac{1}{\alpha \eta} \frac{u(M)}{M \theta} g(M) \right)^2 - \theta \left( g(M) + \frac{u^{\prime}(M)}{u(M)} \frac{M \theta}{u(M)} \right)^2. \label{ODE for u}
\end{equation}
Regarding the boundary conditions, since both $\underline{M}$ and $\overline{M}$ are endogenous and remain to be determined, we apply a change of variables by setting:
\begin{equation}
    M(z) = \underline{M} + \Delta M \cdot z, 
    \quad \text{where} \quad \Delta M = \overline{M} - \underline{M}, \quad z \in [0,1]. \notag
\end{equation}
Then, we define a set of auxiliary functions:
\begin{equation}
    y_1(z) = u(M(z)), 
    \quad y_2(z) = y_1^{\prime}(z) = \Delta M \cdot u^{\prime}(M), 
    \quad y_3(z) = y_2^{\prime}(z) = \Delta M^2 \cdot u^{\prime\prime}(M), \notag
\end{equation}
where $u^{\prime\prime}(M)$ can be represented by $y_1$ and $y_2$, and the boundary conditions are given by: 
\begin{equation}
    y_1(0) = 1 + \gamma, \quad y_1(1) = 1, \quad y_2(0) = y_2(1) = 0. \notag
\end{equation}
This change of variables maps the free-boundary problem with unknown $\underline{M}$ and $\overline{M}$ onto a fixed domain $z \in [0,1]$, which not only standardizes the boundary conditions but also facilitates numerical implementation by allowing the use of standard boundary value problem solvers. 

Regarding the existence and uniqueness of market equilibrium, \citet{henriet2016dynamics} rigorously proved that it holds for the result in Proposition \ref{Proposition Benchmark Equilibrium} when there is no concern for model uncertainty. In contrast, for Proposition \ref{Proposition Robust Equilibrium}, both the structure of the equation and the boundary conditions are altered, which prevents us from providing a rigorous proof of existence and uniqueness. 

Nevertheless, our numerical analysis in Section \ref{Section Numerical} shows that, under standard parameter specifications, the ODE \eqref{ODE for u} with the specified boundary conditions indeed admits a solution satisfying Assumption \ref{Assumption market-to-book ratio}, namely boundedness and monotonicity. Moreover, there exist positive values $\underline{M}^{\ast}$ and $\overline{M}^{\ast}$ that satisfy the boundary conditions. Given one of the boundaries $\underline{M}^{\ast}$ (or $\overline{M}^{\ast}$) fixed, if the ODE \eqref{ODE for u} admits a solution satisfying Assumption \ref{Assumption market-to-book ratio}, then this solution must be unique. The boundedness of $u^{\ast}(\cdot)$ guarantees that the right-hand side of \eqref{ODE for u} is locally Lipschitz in $(u, u^{\prime})$, which ensures uniqueness of the solution.\footnote{Establishing uniqueness in the presence of free boundaries would be more challenging, and we do not pursue this direction further, as it is not the main focus of the present paper.}

\subsection{Properties}

Next, we discuss some properties of the equilibrium. In the benchmark model of \citet{henriet2016dynamics}, it can be rigorously proven that $u^{\ast \prime}(M) \leq 0$, implying that the market-to-book ratio of the insurance sector is a monotonically decreasing function of aggregate capacity. This property not only guarantees the existence of a barrier-type equilibrium, but also suggests that insurers’ expected underwriting profits under the physical measure are non-negative; see Proposition \ref{Proposition Benchmark Properties}. 

In \citet{luciano2022fluctuations}, it is further emphasized that the ratio $R(M) \triangleq -\frac{u^{\prime}(M)}{u(M)}$, can be interpreted as the coefficient of market (absolute) risk aversion, which governs insurers’ underwriting and investment behavior and gives rise to insurance cycles. Since $u^{\ast \prime}(M) \leq 0$, it follows that $R^{\ast}(M) \geq 0$, indicating that insurers behave as if they were risk averse, even though shareholders are risk neutral when discounting future cash flows. In the equilibrium with concerns for model uncertainty, we obtain similar properties. 

\begin{Proposition} \label{Proposition Expected Profit}
    In the equilibrium characterized in Proposition \ref{Proposition Robust Equilibrium}, and under Assumption \ref{Assumption market-to-book ratio} which ensures $u^{\ast \prime}(M) \leq 0$, insurers behave as if they were risk averse, and their expected underwriting profit at the equilibrium price under the optimally chosen measure $\mathbb{Q}^{h^{\ast}}$ is non-negative.
\end{Proposition}

Under the optimally chosen measure, the expected profit per unit of underwritten insurance is given by: 
\begin{equation}
    \eta (p^{\ast}_t + h^{\ast}_t) 
    = - \frac{u^{\ast \prime}(M_t)}{u^{\ast}(M_t)} D(p^{\ast}(M_t)) \eta^2 
    = R^{\ast}(M_t) D(p^{\ast}(M_t)) \eta^2. \label{effective market price}
\end{equation}
This expression consists of two components. The term $p^{\ast}_t$ represents the market price of insurance, while the density generator $h^{\ast}_t$ captures the market price of model uncertainty \citep{anderson2003quartet}. Their sum, $p^{\ast}_t + h^{\ast}_t$, therefore reflects the effective market price of underwriting per unit of risk in the presence of model uncertainty. In particular, $h^{\ast}_t$ can be interpreted as the additional premium required to compensate insurers for the risk of model misspecification, thereby serving as the shadow price of robustness. When $h^{\ast}_t = 0$, the pricing rule of insurance collapses to the benchmark case characterized in Proposition \ref{Proposition Benchmark Equilibrium}. 

The expected profit per unit of underwritten risk is non-negative if and only if $u^{\ast \prime}(M) \leq 0$, which is equivalent to insurers behaving as if they were risk averse. Moreover, the ratio of the effective market price to market demand (which also equals total supply), representing the effective profit margin, is proportional to the market risk-aversion level. This implies that insurers with greater market risk aversion require higher profit margins as compensation for bearing market risk. 

In \citet{luciano2022fluctuations}, it is further shown that market risk aversion $R(M)$ decreases with aggregate capacity, meaning that larger liquid reserves make insurers appear less risk averse. However, this monotonic property does not carry over to the robust equilibrium, since the presence of model uncertainty alters insurers’ effective pricing of risk. In particular, at the boundaries, dividend distributions and external financing dominate underwriting decisions, and the no-arbitrage condition implies that insurers are effectively neutral to incremental risk-taking. We will examine the dynamics of market risk aversion in greater detail in Section \ref{Section Numerical}. 

\begin{Proposition} \label{Proposition Limiting theta}
    In the limit as $\theta \rightarrow \infty$, insurers effectively fully trust the physical probability model, thereby eliminating concerns about model uncertainty. Consequently, the system of equations that $p^{\ast}(\cdot)$ and $u^{\ast}(\cdot)$ must satisfy converges to the benchmark case without model uncertainty, although the boundary conditions remain different.
\end{Proposition} 

Mathematically, the market-clearing condition leads to: 
\begin{equation}
    \frac{M}{u^{\ast}(M)\eta} \left( p^{\ast}(M) + \frac{u^{\ast \prime}(M)}{u^{\ast}(M)} D(p^{\ast}(M)) \eta \right) = \frac{1}{\theta} \left( 1 + M \frac{u^{\ast \prime}(M)}{u^{\ast}(M)} \right) D(p^{\ast}(M)) \xrightarrow[]{\theta \rightarrow \infty} 0, \notag 
\end{equation}
which implies $p^{\ast}(M) + \frac{u^{\ast \prime}(M)}{u^{\ast}(M)} D(p^{\ast}(M)) \eta \rightarrow 0$, if $M \geq \varepsilon > 0$, where $\varepsilon$ is a small fixed constant ensuring that aggregate reserves are strictly positive. This result coincides with the first equation in \eqref{Solution to Benchmark Equilibrium} and further implies that the optimal density generator satisfies $h^{\ast}(M) \rightarrow 0$, indicating that insurers no longer distort probability measures in the limit of vanishing model uncertainty. Similarly, given $M \geq \varepsilon > 0$, the second equilibrium condition becomes:
\begin{align}
    2r & = \left( p^{\ast}(M) + \frac{u^{\ast \prime}(M)}{u^{\ast}(M)} D(p^{\ast}(M)) \eta \right)\left(\frac{1}{M} + \frac{u^{\ast \prime}(M)}{u^{\ast}(M)} \right) D(p^{\ast}(M)) \eta \notag \\
    & + \left[\frac{u^{\ast \prime \prime}(M)}{u^{\ast}(M)} - 2{\left(\frac{u^{\ast \prime}(M)}{u^{\ast}(M)}\right)}^2\right] D(p^{\ast}(M))^2 \eta^2 \notag \\
    & \xrightarrow[]{\theta \rightarrow \infty} \left[\frac{u^{\ast \prime \prime}(M)}{u^{\ast}(M)} - 2{\left(\frac{u^{\ast \prime}(M)}{u^{\ast}(M)}\right)}^2\right] D(p^{\ast}(M))^2 \eta^2, \notag  
\end{align}
which is consistent with the second equation in \eqref{Solution to Benchmark Equilibrium}. Therefore, as insurers’ confidence in the physical probability model becomes arbitrarily large, the ODE satisfied by the market-to-book ratio formally converges to the benchmark case without model uncertainty on the domain $M \geq \varepsilon$. However, this convergence does not extend to the region near $M \rightarrow 0$, and the boundary condition at $\underline{M}$ remains different from the benchmark case.

In our numerical analysis, we show that as $\theta$ becomes sufficiently large, the solved value of $\underline{M}$ approaches zero, although it does not exactly coincide with the benchmark boundary. Moreover, both the market-to-book ratio and the price function converge closely to their benchmark counterparts, with the corresponding curves almost overlapping.

\section{Quantitative Analysis} \label{Section Numerical}

In this section, we conduct numerical simulations to solve the ODE system, characterize the dynamic equilibrium of the insurance market, and visually examine the impact of insurers' concerns about model uncertainty. To facilitate comparison, we define $p^{\ast}_b$, $u^{\ast}_b$ as the benchmark equilibrium solutions without model uncertainty; and $p^{\ast}_r$, $u^{\ast}_r$ as the equilibrium under robustness. 

As the benchmark setting, we specify the parameter values as follows. The risk-free interest rate is set to $r = 0.04$, corresponding to a 4\% annualized rate, which serves as the discount factor for shareholders’ valuation of future cash flows. Following the estimation of \citet{luciano2022fluctuations}, we set the unit loss to $l = 1.0$ and the volatility of unit loss risk to $\eta = 0.28$; the cost of external financing is parameterized as $\gamma = 0.20$, interpreted as the expected return on equity financing; the risk aversion coefficient of insurees (general households) is set to $\alpha = 2.0$. Finally, referring to \citet{ling2021robust}, we adopt $\theta = 2.8$ as the benchmark degree of robustness concern. 

In numerical computation, We employ MATLAB’s bvp4c (a fourth-order Lobatto collocation method with adaptive meshing) and treat the free boundaries $\underline{M}$ and $\overline{M}$ as unknown parameters, solving them jointly with the states from a smooth initial guess until convergence.

\begin{table}[htbp]
\centering
\setlength{\tabcolsep}{20pt}
\renewcommand{\arraystretch}{1.4}
\small
\begin{tabular}{lcc}
\hline
Parameter & Symbol & Value \\
\hline
Risk-free rate                & $r$       & 4\% \\
Expectation of loss risk       & $l$    & 1.0 \\
Volatility of loss risk       & $\eta$    & 28\% \\
Cost of external financing    & $\gamma$  & 20\% \\
Risk aversion of insuree      & $\alpha$  & 2.0 \\
Robustness parameter          & $\theta$  & 2.8 \\
\hline
\end{tabular}
\caption{Benchmark Parameter Values.}
\end{table}

\subsection{Comparison of Equilibria with and without Model Uncertainty} \label{Subsection Comparison of Equilibria}

\begin{figure}[t]
    \centering
    \includegraphics[width=0.45\linewidth]{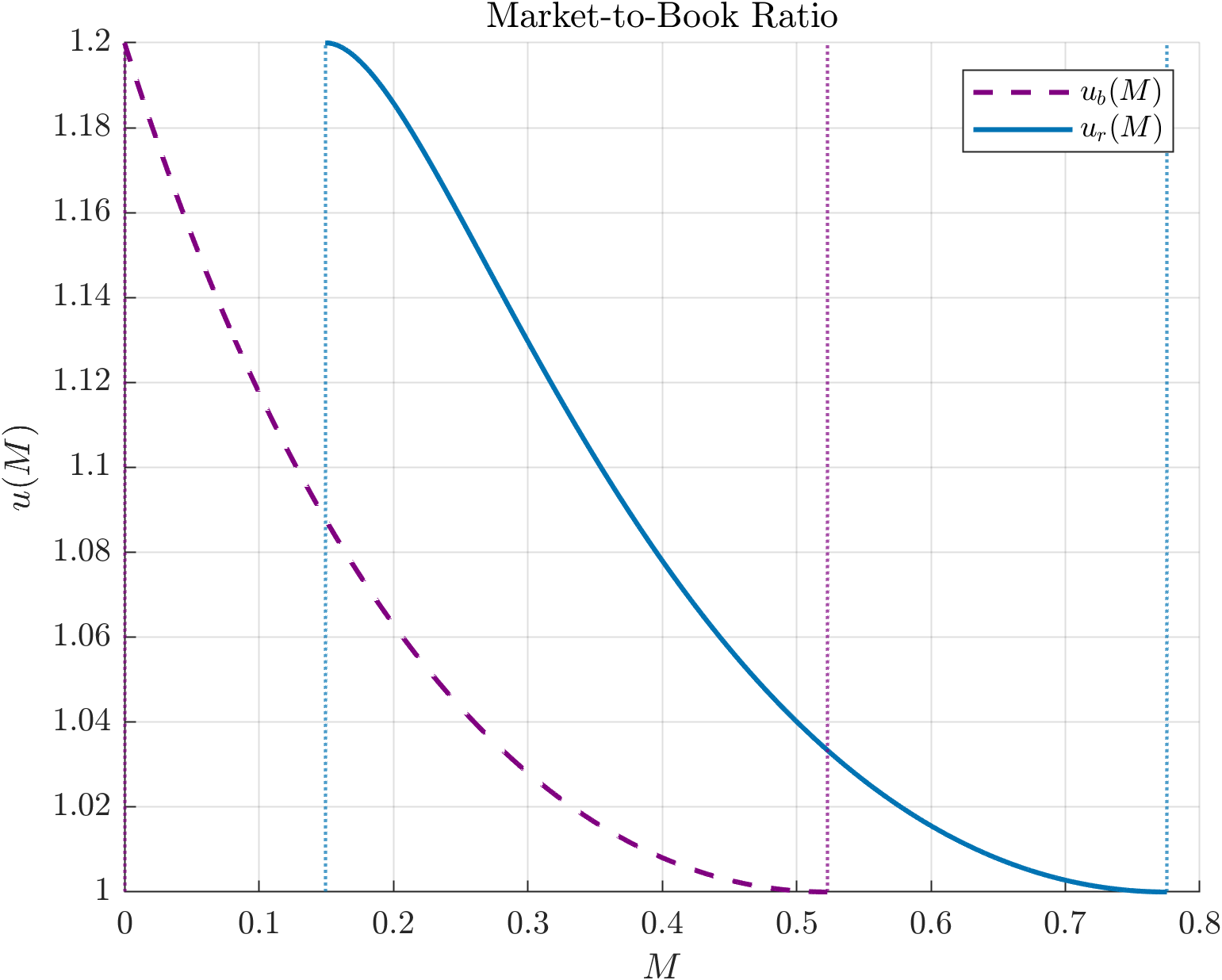}
    \includegraphics[width=0.45\linewidth]{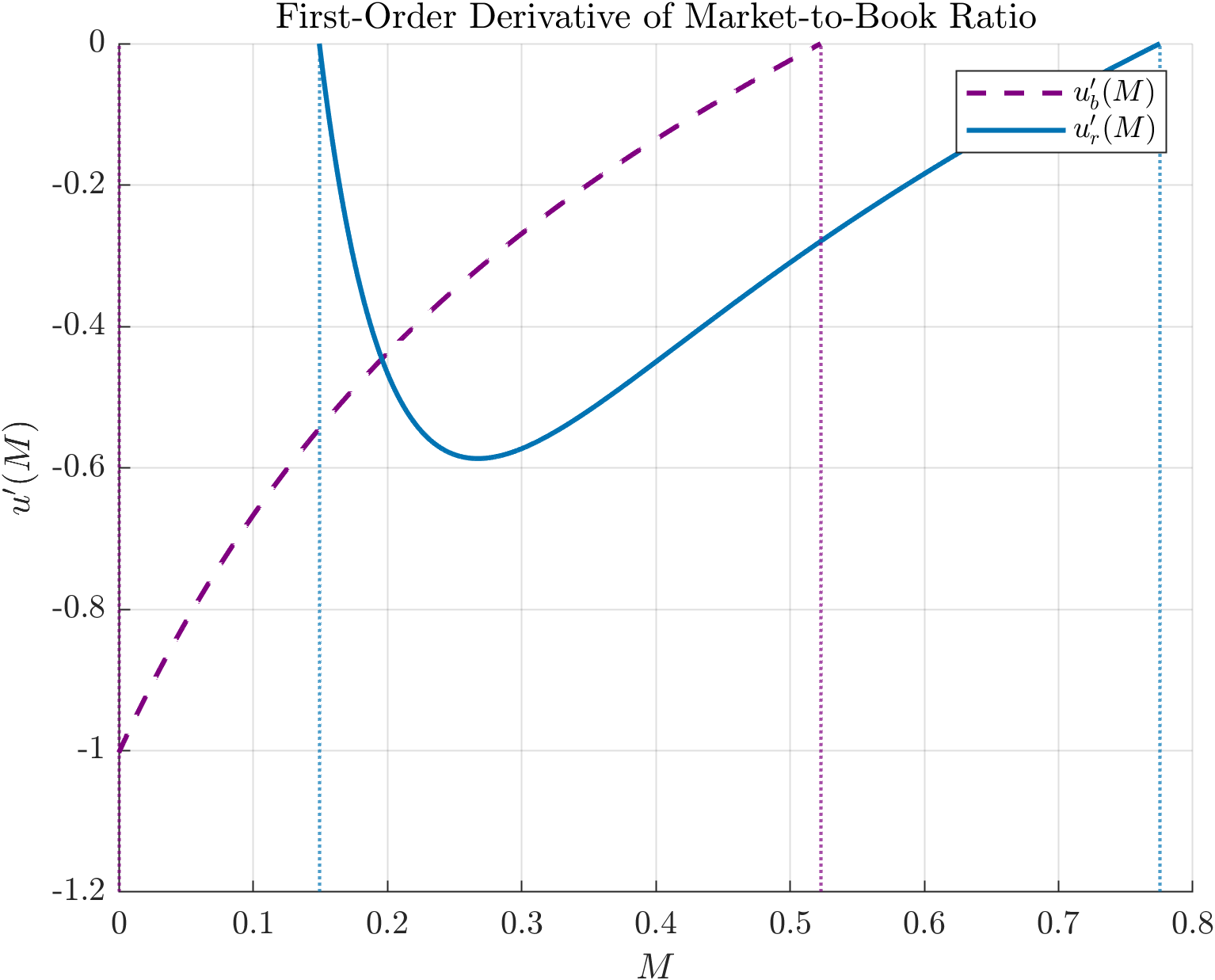}
    \includegraphics[width=0.45\linewidth]{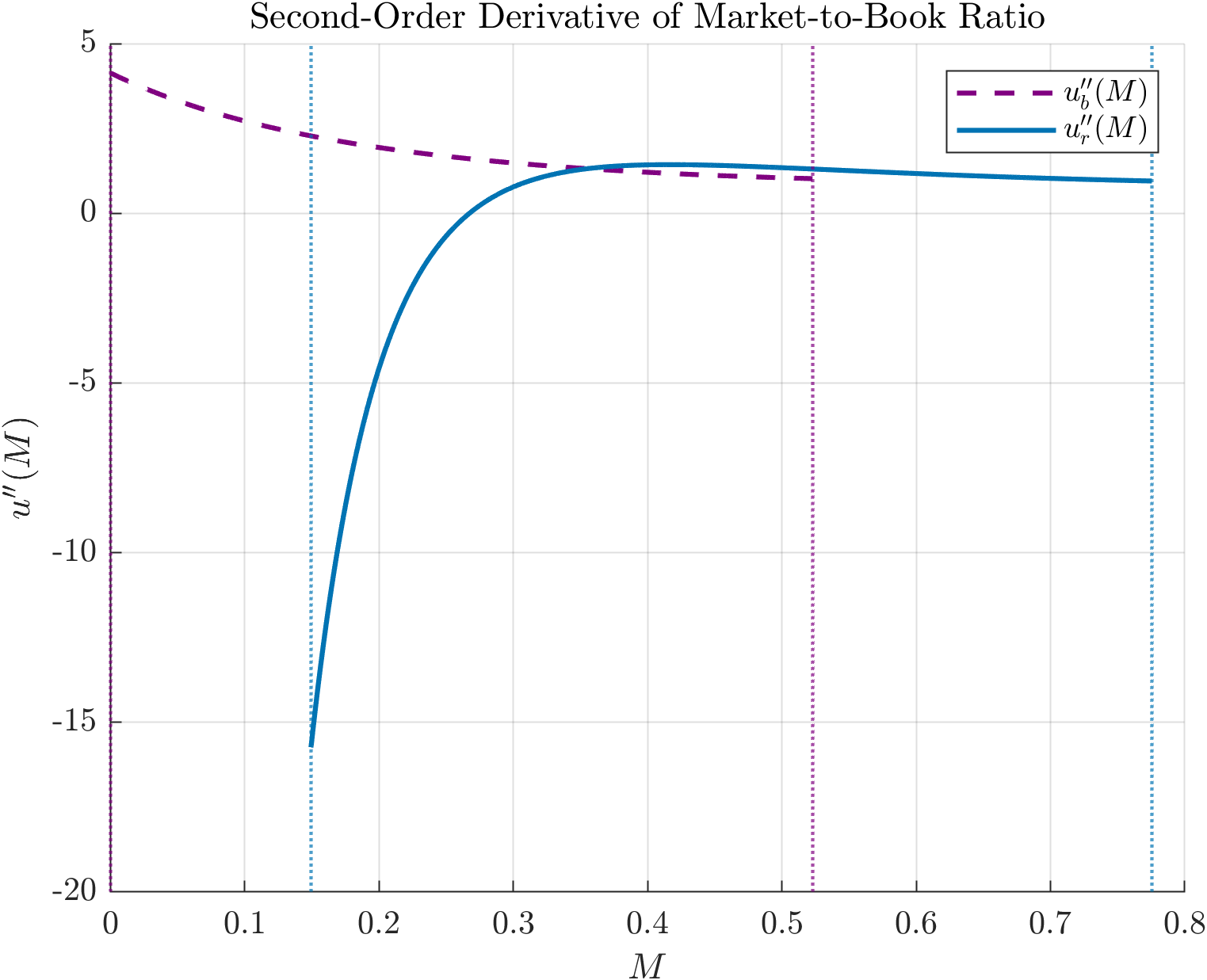}
    \includegraphics[width=0.45\linewidth]{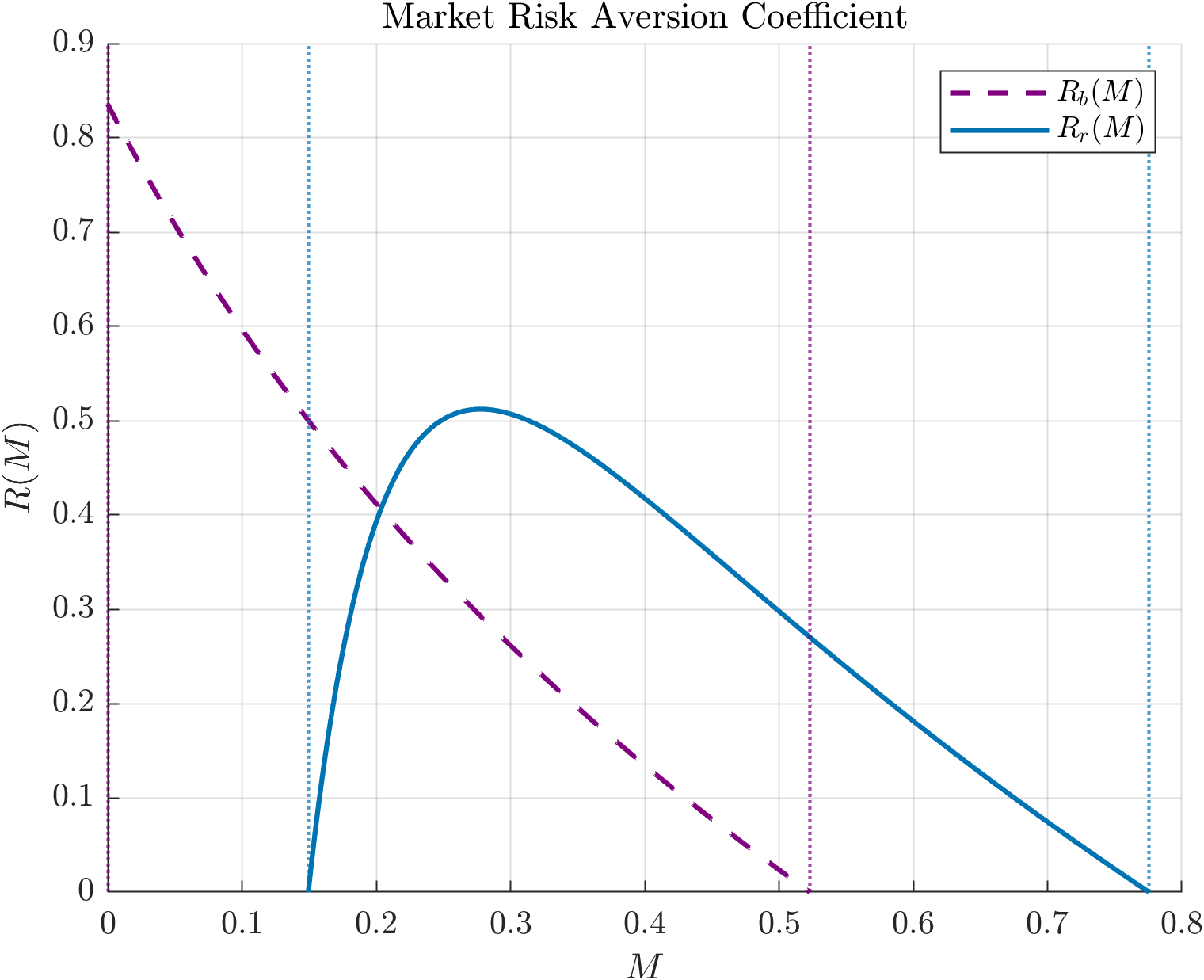}
    \caption{Equilibrium Market-to-Book Ratio with and without Model Uncertainty.}
    \label{Figure Comparison of Market-to-Book Ratio}
\end{figure}

The equilibria characterized by Propositions \ref{Proposition Benchmark Equilibrium} and \ref{Proposition Robust Equilibrium} can be numerically solved without difficulty. Figure \ref{Figure Comparison of Market-to-Book Ratio} reports the results for the market-to-book ratio, its first- and second-order derivatives, and the implied market risk aversion $R(M)$. The solid blue lines correspond to the case with concerns about model uncertainty, while the dashed purple lines represent the benchmark case without model uncertainty. Several notable features can be observed from the comparison. 

First, the minimum, maximum, and range of aggregate capacity all expand when accounting for model uncertainty. In the benchmark case, $\underline{M}_{b} = 0$, $\overline{M}_{b} = 0.52$, and $\Delta M_b = 0.52$, whereas in the robust case, $\underline{M}_{r} = 0.15$, $\overline{M}_{r} = 0.78$, and $\Delta M_r = 0.63$. The higher lower bound indicates that insurers under model uncertainty must hold larger minimum reserves, reflecting a more conservative stance. The higher upper bound suggests that the market can sustain a greater accumulation of reserves before dividend payouts occur. Consequently, the wider interval between the two barriers reflects more cautious capital management overall: dividend distributions are postponed, while the trigger for external recapitalization is also elevated.

Second, at the same aggregate capacity level, the market-to-book ratio in the robust equilibrium becomes significantly higher. Economically, this reflects that insurers demand a higher valuation of equity, as they require additional compensation for bearing the risk of model misspecification. In other words, robustness concerns raise the shadow cost of capital. While the ratio remains monotonically decreasing, it is no longer convex: the marginal decline in valuation first accelerates and then decelerates, reflecting a change in the curvature of equity valuation.

Third, the implied market risk aversion is no longer decreasing but instead increases initially and then decreases. This non-monotonicity arises because robustness concerns amplify risk perceptions differently across capacity levels. At low reserves, external recapitalization dominates, so effective risk aversion starts from zero. At intermediate reserves, the fear of ambiguous risk is most pronounced, raising insurers’ required compensation. At high reserves, abundant capital buffers mitigate the marginal effect of ambiguity, causing risk attitudes to gradually revert toward neutrality.

\begin{figure}[t]
    \centering
    \includegraphics[width=0.45\linewidth]{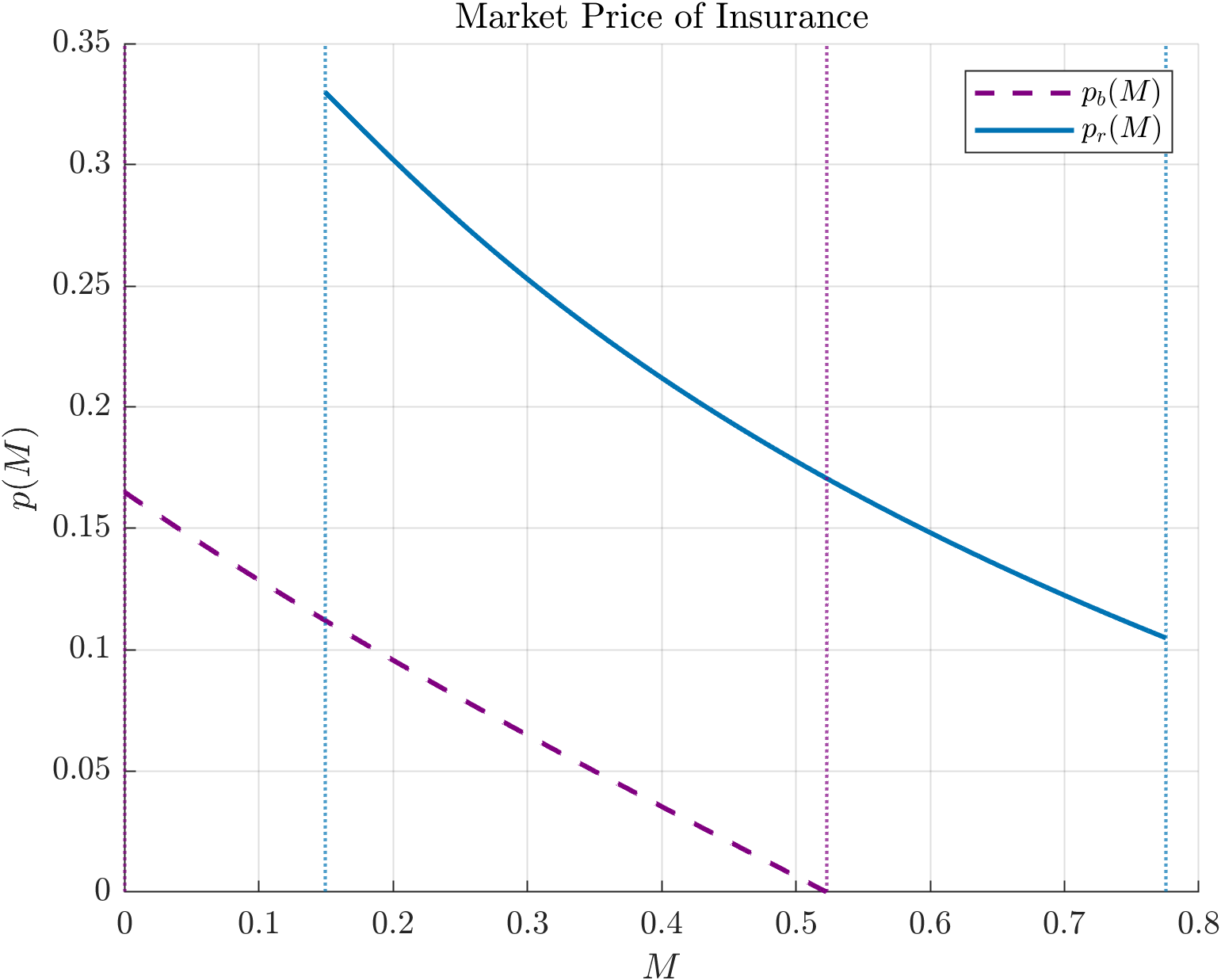}
    \includegraphics[width=0.45\linewidth]{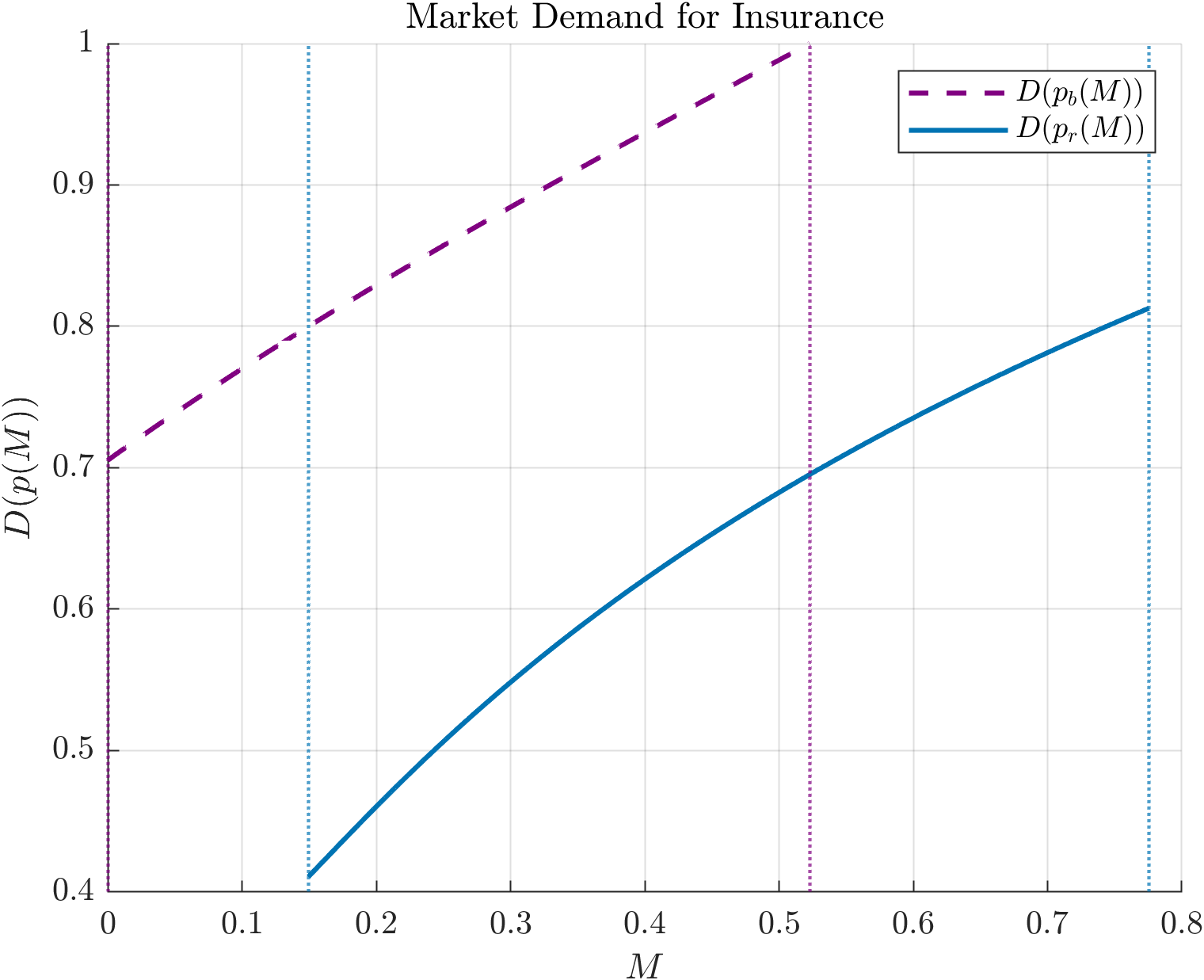}
    \includegraphics[width=0.45\linewidth]{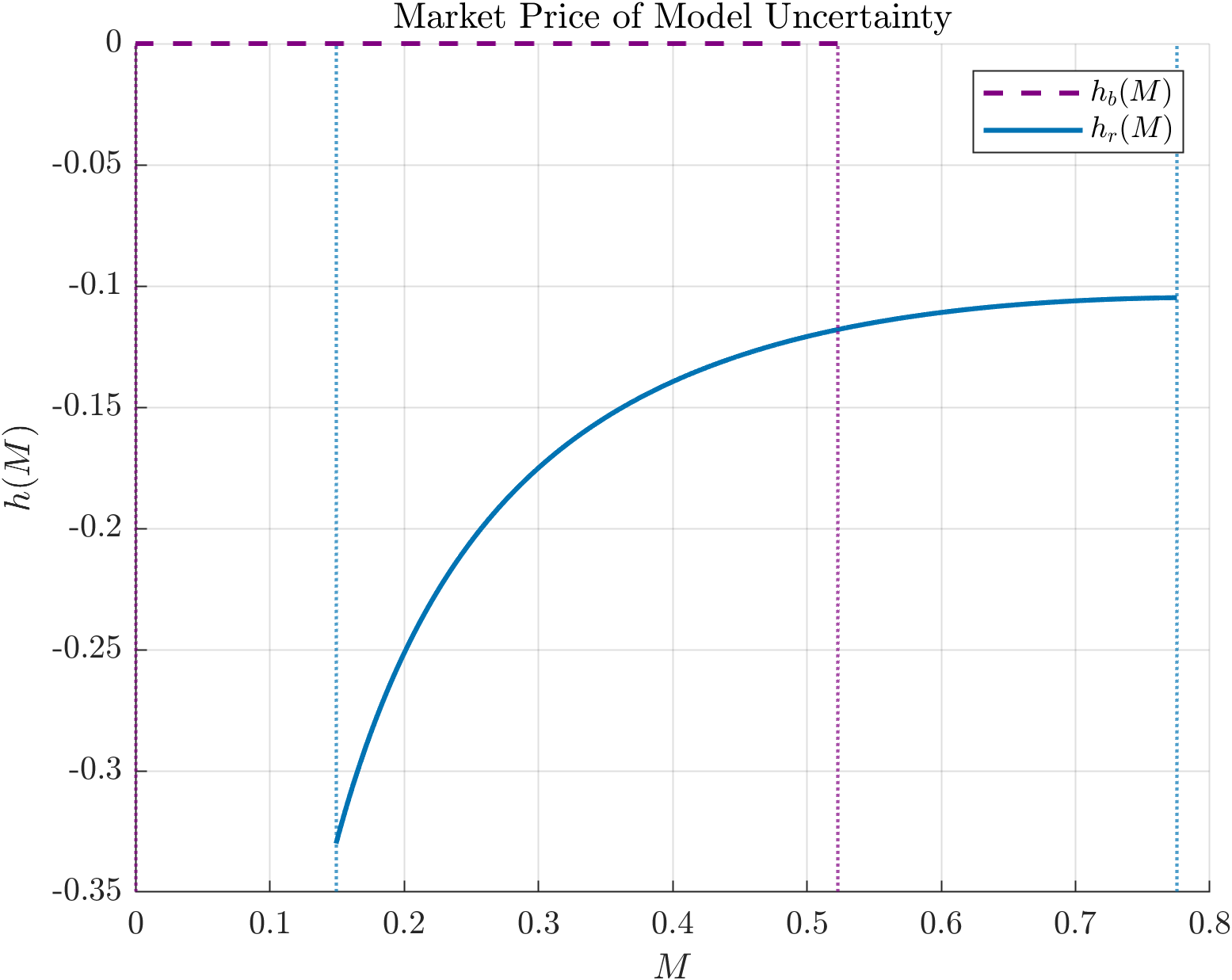}
    \includegraphics[width=0.45\linewidth]{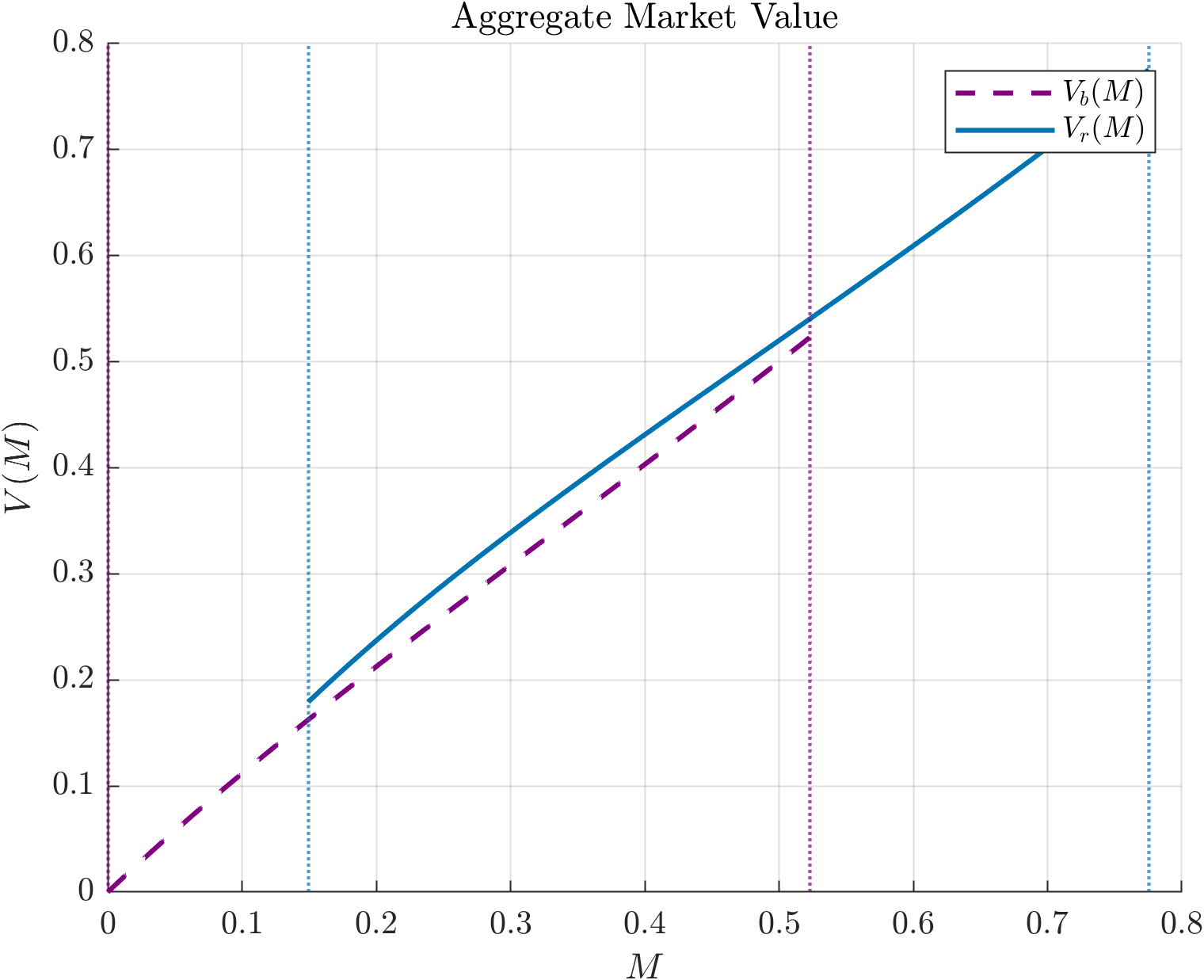}
    \caption{Equilibrium Insurance Market with and without Model Uncertainty.}
    \label{Figure Comparison of Insurance Market}
\end{figure}

Next, we analyze the differences in insurance market outcomes induced by model uncertainty, as shown in Figure \ref{Figure Comparison of Insurance Market}. 
Consistent with the intuition that insurers behave more conservatively, the equilibrium price of insurance is significantly higher for the same level of capacity. Quantitatively, the price increases by about 0.15-0.20 units, corresponding to a 4.2\%-5.6\% rise in the premium per unit of risk. While the equilibrium price remains a decreasing function of aggregate capacity, market demand correspondingly increases, highlighting the role of capital adequacy in insurance pricing. The market price of model uncertainty also rises with capacity and gradually approaches zero, indicating that concerns about model misspecification diminish when reserves are sufficiently abundant. Finally, the aggregate market value is higher under the robust equilibrium, suggesting that robustness concerns, while inducing more conservative behavior at the firm level, enhance the overall value of the insurance sector.

\subsection{Impact of Robustness Degree}

\begin{figure}[t]
    \centering
    \includegraphics[width=0.45\linewidth]{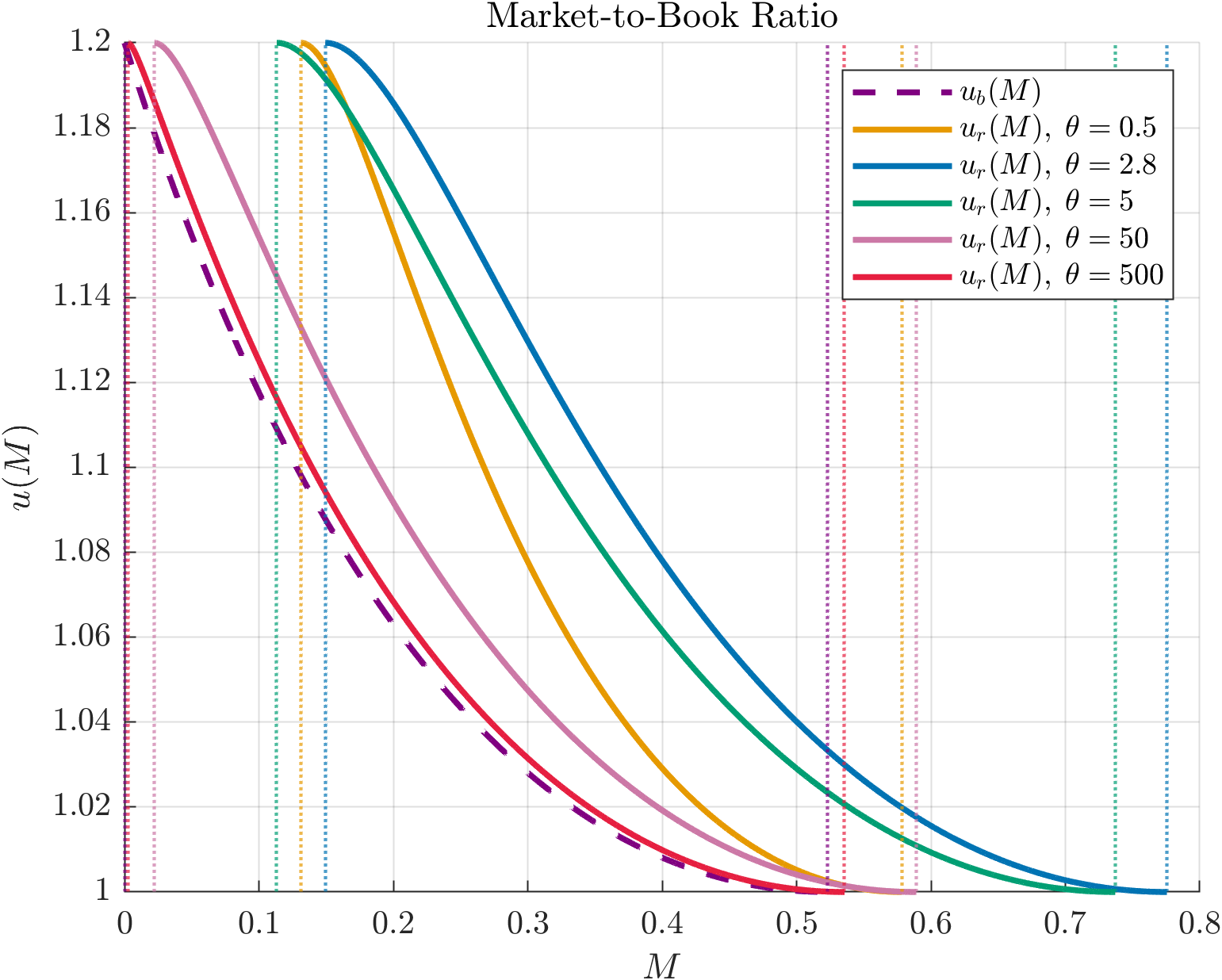}
    \includegraphics[width=0.45\linewidth]{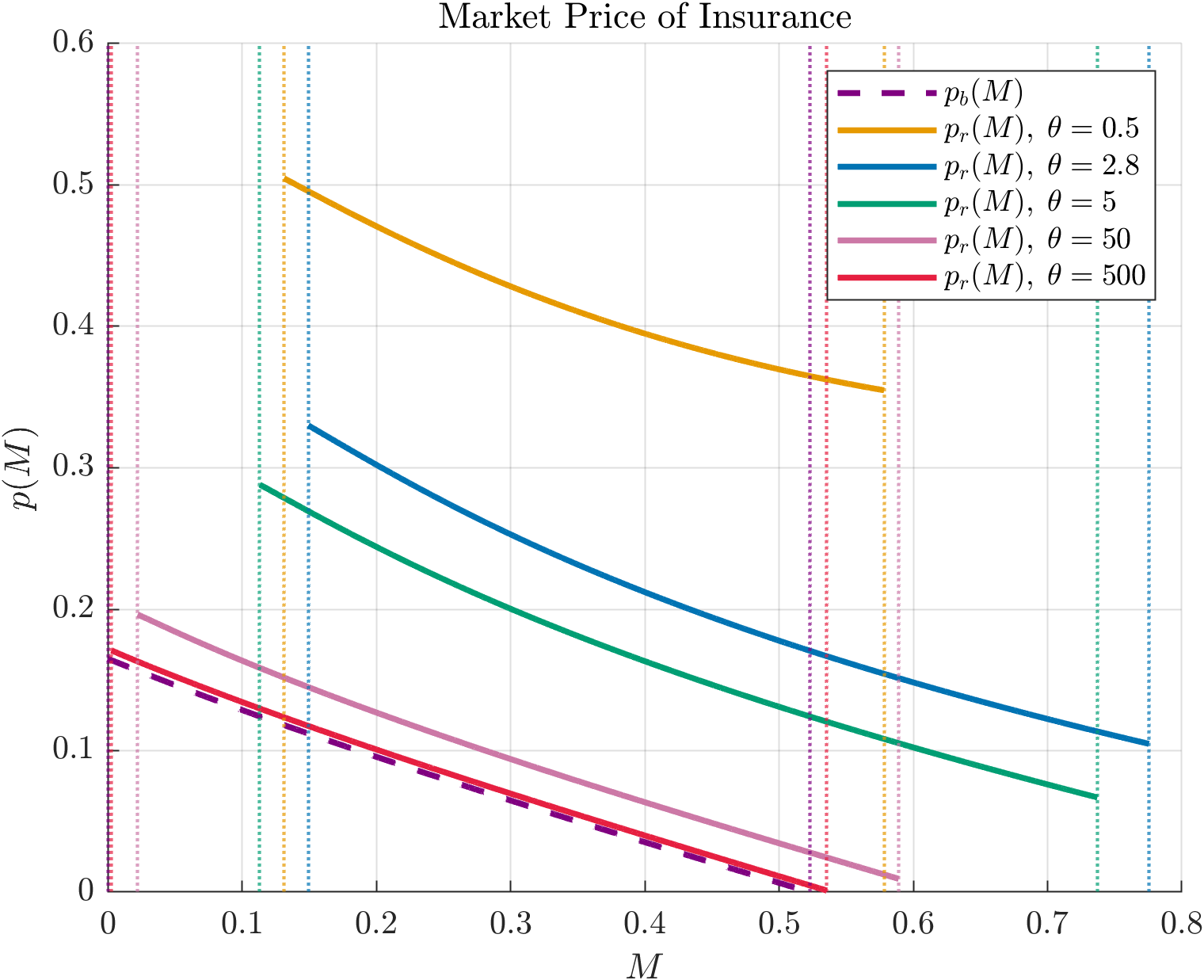}
    \caption{Equilibrium Outcomes for Different Robustness Degree $\theta$.}
    \label{Figure Different Robustness Degree}
\end{figure}

We next study how the degree of concern for robustness affects the equilibrium outcomes. Theoretically, a larger $\theta$ corresponds to lower ambiguity aversion, meaning that insurers place greater trust in the reference probability model. Proposition \ref{Proposition Limiting theta} suggests that as $\theta \to \infty$, the ODE system solved by $p_r^{\ast}(\cdot)$ and $u_r^{\ast}(\cdot)$ converges to that of the benchmark case $p_b^{\ast}(\cdot)$ and $u_b^{\ast}(\cdot)$, although the boundary conditions remain different. Figure \ref{Figure Different Robustness Degree} illustrates this convergence by plotting the equilibrium market-to-book ratio and the market price of insurance for values of $\theta$ ranging from $0.5$ to $500$. 
When $\theta = 500$, the equilibrium curves (solid red curves) almost coincide with those of the benchmark case (purple dashed curves), confirming the limiting result. Consistently, the external financing and payout boundaries, $\underline{M}_{r, \theta = 500} = 0.0027$ and $\overline{M}_{r, \theta = 500} = 0.5356$, are also very close to their benchmark counterparts, $\underline{M}_{b} = 0$ and $\overline{M}_b = 0.52$. 

Turning to comparative statics, there seems to be a clear monotonic relationship between the robustness degree and insurance prices. 
Given the capacity level, the equilibrium price decreases as $\theta$ increases, consistent with the intuition that more ambiguity-averse insurers (smaller $\theta$) require higher premia as compensation for model misspecification. 

\begin{table}[htbp]
\centering
\setlength{\tabcolsep}{8pt}
\renewcommand{\arraystretch}{1.4}
\small
\begin{tabular}{cccc}
\hline
Parameter       & \multicolumn{1}{c}{External Financing Boundary, $\underline{M}$} & Payout Boundary, $\overline{M}$ & Range of Capacity, $\Delta{M}$ \\ \hline
$\theta = 0.5$  & 0.1313                                                            & 0.5783                          & 0.4470                         \\
$\theta = 2.8$  & 0.1494                                                            & 0.7755                          & 0.6261                         \\
$\theta = 5$    & 0.1133                                                            & 0.7371                          & 0.6238                         \\
$\theta = 50$   & 0.0222                                                            & 0.5891                          & 0.5669                         \\
$\theta = 500$  & 0.0027                                                            & 0.5356                          & 0.5329                         \\ \hline
$\gamma = 0.06$ & 0.2147                                                            & 0.6609                          & 0.4462                         \\
$\gamma = 0.1$  & 0.1876                                                            & 0.7060                          & 0.5184                         \\
$\gamma = 0.2$  & 0.1494                                                            & 0.7755                          & 0.6261                         \\
$\gamma = 0.3$  & 0.1273                                                            & 0.8188                          & 0.6915                         \\
$\gamma = 0.4$  & 0.1120                                                            & 0.8493                          & 0.7373                         \\ \hline
\end{tabular}
\caption{Numerically Solved Boundary Values.}
\label{Table Boundary Values}
\end{table}

In contrast, no such monotonicity holds for the market-to-book ratio. For instance, the curve with $\theta = 0.5$ lies between other curves, indicating that at the same capacity level, the market-to-book ratio first increases and then decreases with $\theta$. A similar non-monotonic pattern also emerges for the external financing and payout boundaries. As summarized in Table \ref{Table Boundary Values}, both boundaries and the resulting range of admissible capacity first expand with $\theta$ and then contract, suggesting that robustness concerns affect liquidity management in a non-linear fashion. Nevertheless, they remain consistently above the benchmark case, confirming the robustness of our analysis in Subsection \ref{Subsection Comparison of Equilibria}.

\subsection{Impact of Financing Cost}

Further, we examine the impact of external financing cost $\gamma$ on equilibrium outcomes, which reflects the degree of financial friction faced by insurers. It can also be interpreted as the required long-term rate of return demanded by capital providers for investing in the insurance industry. 

\begin{figure}[t]
    \centering
    \includegraphics[width=0.45\linewidth]{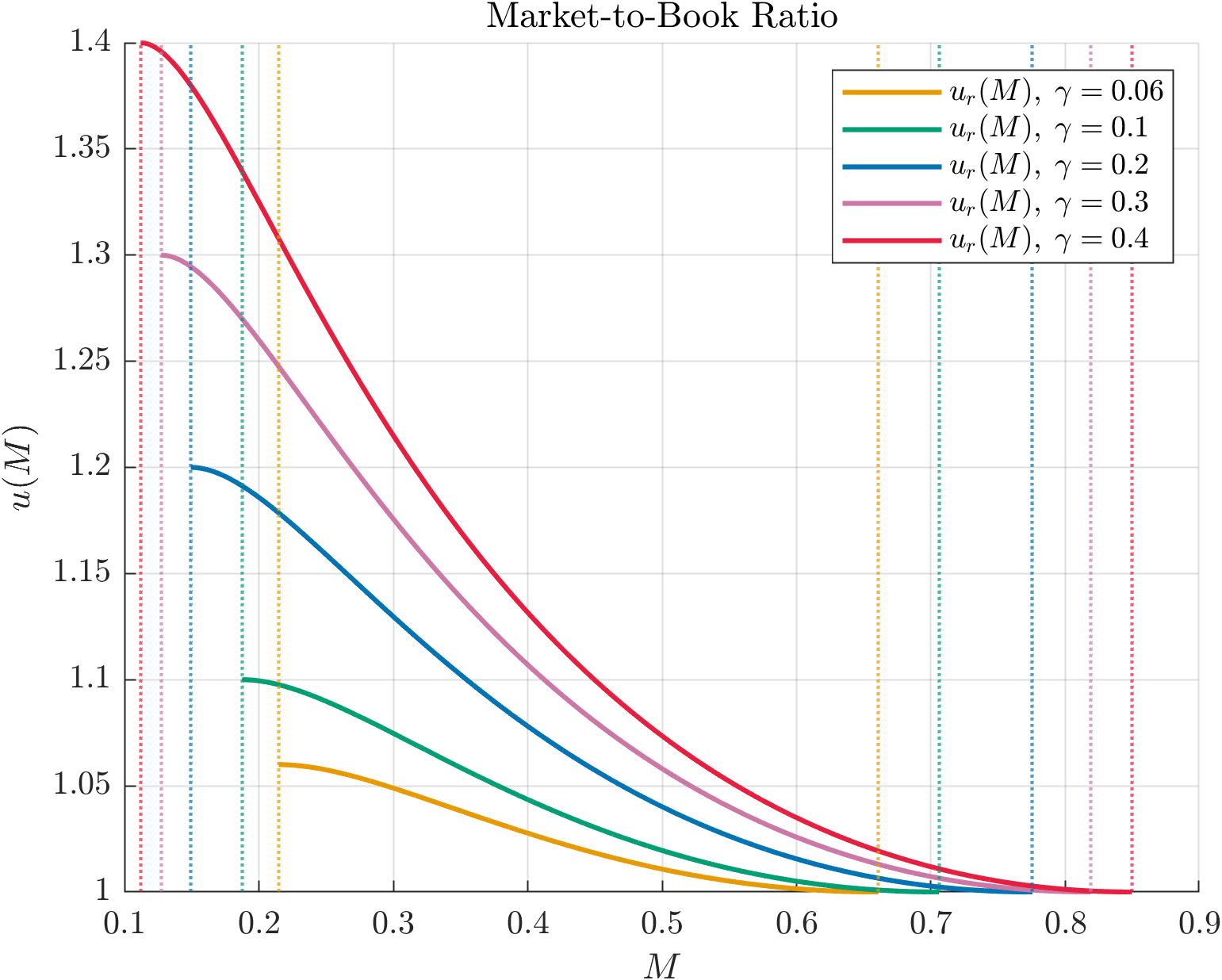}
    \includegraphics[width=0.45\linewidth]{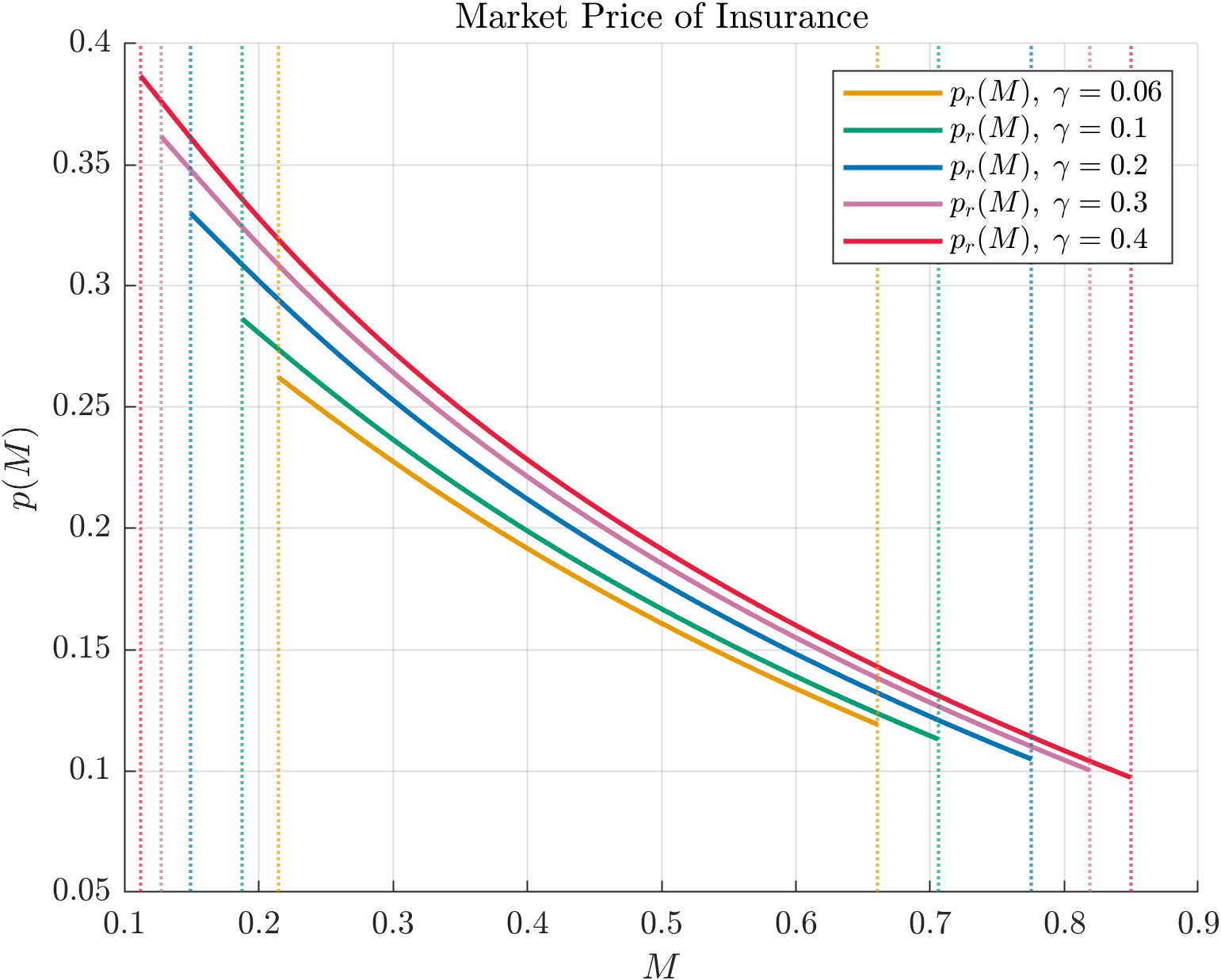}
    \caption{Equilibrium Outcomes for Different Financing Cost $\gamma$.}
    \label{Figure Different Financing Cost}
\end{figure}

First, as illustrated in Figure \ref{Figure Different Financing Cost}, we observe that as the cost $\gamma$ rises from $6\%$ to $40\%$, the equilibrium market price of insurance increases monotonically. This relationship also appears in the benchmark equilibrium of \citet{henriet2016dynamics} without model uncertainty. Intuitively, when the required return on external capital is higher, or equivalently when the cost of funds is higher, insurers charge higher premia to compensate for the more expensive capital.  

Second, the market-to-book ratio also exhibits a monotonic increasing relationship with $\gamma$. Economically, a higher financing cost makes equity capital more valuable, since shareholders demand greater compensation for supplying funds. This raises the valuation of each unit of equity relative to book value, leading to a higher market-to-book ratio.  

Third, as reported in Table \ref{Table Boundary Values}, both boundaries and the admissible range of aggregate capacity expand with $\gamma$. A higher financing cost makes external recapitalization more expensive, forcing insurers to maintain larger precautionary reserves (higher lower bound) and to delay dividend payouts (higher upper bound). The wider interval thus reflects more conservative liquidity management. These results also highlight that the amplitude of underwriting cycles is intrinsically linked to the severity of financial frictions in the insurance market.

\section{Insurers’ Long-Run Behavior Pattern} \label{Section Long-Run Behavior}

In this section, we simulate the insurance market equilibrium in a dynamic setting to gain deeper insights into the long-run behavior of insurers. We continue to rely on the benchmark parameter specifications introduced in Section \ref{Section Numerical}, together with the corresponding numerical solutions of the equilibrium. We then investigate how variations in key parameters shape the behavior patterns.

\subsection{Underwriting Dynamics}

Under the optimally chosen underwriting and pricing strategies, the aggregate capacity of the insurance sector evolves as follows: 
\begin{align}
    \mathrm{d}M_{b, t} & = D\!\left(p_b^{\ast}(M_{b,t})\right) \eta p_b^{\ast}(M_{b,t}) \,\mathrm{d}t 
    - D\!\left(p_b^{\ast}(M_{b,t})\right) \eta \,\mathrm{d}B_t \notag \\
    & = \underbrace{R_b(M_{b,t}) \Big(D(p_b^{\ast}(M_{b,t})) \eta \Big)^2 }_{\mu_b(M)}  \,\mathrm{d}t 
    \underbrace{- D(p_b^{\ast}(M_{b,t})) \eta}_{\sigma_b(M)} \,\mathrm{d}B_t,  
    \quad M_{b,t} \in [0, \overline{M}_b], 
    \label{Equilibrium M Dynamics Without} \\[1em] 
    \mathrm{d}M_{r, t} & = D\!\left(p_r^{\ast}(M_{r,t})\right) \eta \Big(p_r^{\ast}(M_{r,t}) + h^{\ast}(M_{r,t}) \Big) \,\mathrm{d}t 
    -  D\!\left(p_r^{\ast}(M_{r,t})\right) \eta \,\mathrm{d}B_t^{h^{\ast}} \notag \\
    & = \underbrace{R_r(M_{r,t}) \Big(D(p_r^{\ast}(M_{r,t})) \eta \Big)^2}_{\mu_r(M)} \,\mathrm{d}t 
    \underbrace{-  D(p_r^{\ast}(M_{r,t})) \eta}_{\sigma_r(M)} \,\mathrm{d}B_t^{h^{\ast}}, 
    \quad M_{r,t} \in [\underline{M}_{r}, \overline{M}_r],   
    \label{Equilibrium M Dynamics With}
\end{align}
where the first process describes the dynamics under the physical measure $\mathbb{P}$ in the absence of model uncertainty, while the second corresponds to the dynamics under the optimally chosen worst-case measure $\mathbb{Q}^{h^{\ast}}$ in the presence of model uncertainty. Here, $B_t$ denotes a standard Brownian motion under $\mathbb{P}$, and $B_t^{h^{\ast}}$ the Brownian motion under $\mathbb{Q}^{h^{\ast}}$. For numerical comparison, we impose the same Brownian increments across the two measures, so that differences in dynamics are solely attributable to model uncertainty rather than stochastic noise.

\begin{figure}[t]
    \centering
    \includegraphics[width=0.9\linewidth]{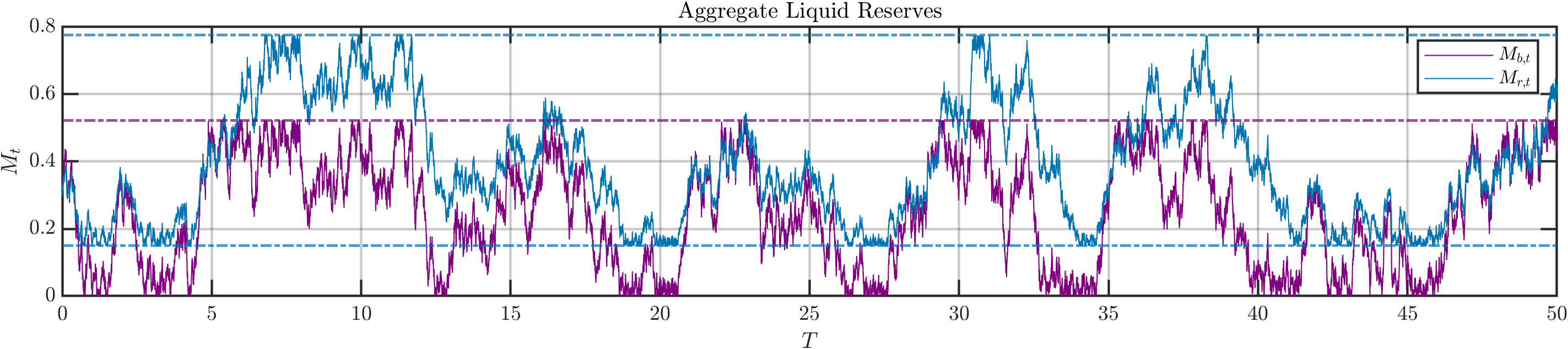}
    \includegraphics[width=0.9\linewidth]{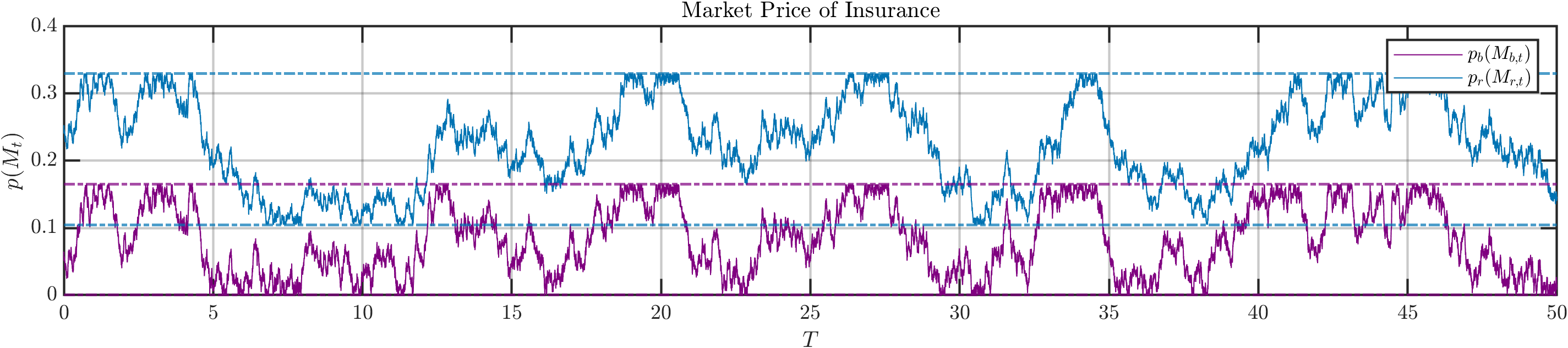}
    \caption{Dynamic Equilibrium Outcomes with and without Model Uncertainty.} 
    \label{Figure Dynamics Comparison}
\end{figure}

Figure \ref{Figure Dynamics Comparison} presents the simulated dynamic equilibrium outcomes for the insurance market, including the reserve process and the price process, with and without model uncertainty. The simulations are initialized with an aggregate capacity of $M_{0} = 0.3$. A key observation is that, regardless of whether model uncertainty is present, the insurance price adjusts dynamically with the level of aggregate capacity, generating the cyclical behavior commonly documented in insurance markets \citep{harrington2013insurance}. The two paths display broadly similar directional movements, which is due to the fact that we impose identical Brownian increments. 

Our primary interest lies in the impact of model uncertainty on market dynamics. Obviously, the robust equilibrium price $p_{r}(M_{r,t})$ is systematically higher than $p_{b}(M_{b,t})$, even though liquid reserves are also maintained at higher levels for most of the time. It reflects the additional premium that insurers require as compensation for model misspecification, while at the same time indicating that capital accumulation in the insurance sector becomes more resilient under robustness concerns. 

A further observation is that the volatility of both the reserve and price processes in the robust equilibrium is relatively smaller than in the benchmark case. As shown in \eqref{Equilibrium M Dynamics Without} and \eqref{Equilibrium M Dynamics With}, the effective diffusion term $\left|\sigma_r(M)\right|$ is generally lower than $\left|\sigma_b(M)\right|$ because of reduced demand. Quantitatively, the drift term $\left|\mu(M)\right|$ is very small compared to $\left|\sigma(M)\right|$, implying that fluctuations within the two boundaries are primarily driven by diffusion. The smaller volatility, combined with a wider admissible capacity range, lengthens the duration of cycles in the robust equilibrium. These features indicate that insurers operate in a more stable and conservative manner under robustness concerns: both reserve flows and price movements exhibit lower volatility, and capital adjustments (recapitalization as well as payouts) occur less frequently, leading to more persistent reserve dynamics.

\begin{figure}[t]
    \centering
    \includegraphics[width=0.9\linewidth]{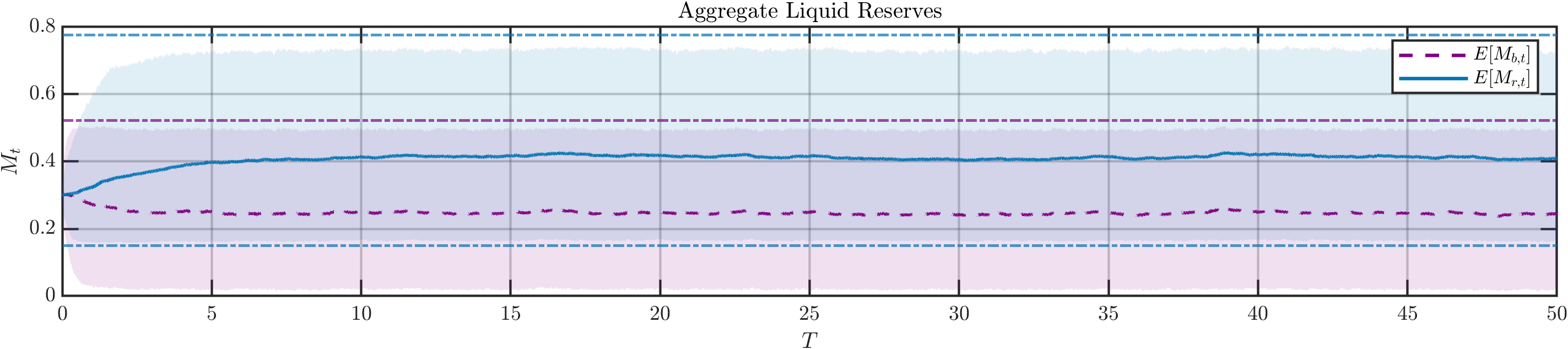}
    \includegraphics[width=0.9\linewidth]{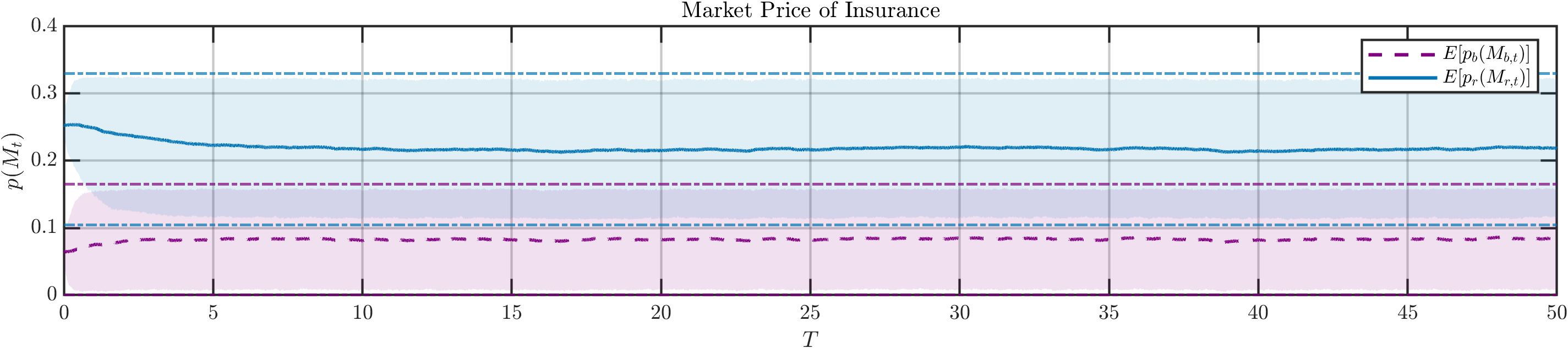}
    \caption{Monte Carlo Simulation of Dynamic Equilibrium Outcomes.} 
    \label{Figure MC Dynamics Comparison}
\end{figure}

Figure \ref{Figure MC Dynamics Comparison} reports the Monte Carlo simulation of the dynamic equilibrium, where 2000 sample paths are generated and the mean together with the 5\%-95\% quantile bands are shown. Both the average aggregate reserves and the average market price are systematically higher in the robust equilibrium than in the benchmark case. As $T$ increases, the mean curves of both equilibria flatten out, indicating that the system converges to a stationary distribution. 

\subsection{Duration of Underwriting Cycles}

We have documented the cyclical nature of insurance capacity and pricing dynamics in the previous analysis. In this subsection, we turn to the impact of model uncertainty on the duration of underwriting cycles. Formally, an underwriting cycle is defined as alternating phases of: (1) a soft market, during which insurers’ aggregate capacity expands from $\underline{M}$ to $\overline{M}$, leading to falling equilibrium prices; and (2) a hard market, during which aggregate capacity contracts from $\overline{M}$ back to $\underline{M}$, accompanied by rising prices. Similar to \citet{henriet2016dynamics}, we compute the expected duration of each phase of the underwriting cycle based on the dynamics in \eqref{Equilibrium M Dynamics Without} and \eqref{Equilibrium M Dynamics With}. 

Formally, let $T_s(M)$ denote the expected time for the reserve process $M_t$ to reach the upper boundary $\overline{M}$ from any state $M \leq \overline{M}$. Then, $T_s(\underline{M})$ captures the expected duration of the soft market phase. Let $T_h(M)$ be the expected time for $M_t$ to return from any state $M \geq \underline{M}$ to the lower boundary $\underline{M}$. Then, $T_h(\overline{M})$ represents the expected duration of the hard market phase. Finally, the total expected duration of an insurance cycle is given by $T_c \triangleq T_s(\underline{M}) +  T_h(\overline{M})$. 

\begin{Proposition}
    Let $o\in\{b,r\}$ index the benchmark and robust equilibria, and let the aggregate capacity $M_{o,t}$ evolve on $[\underline M_o,\overline M_o]$ with drift $\mu_o(M)$ and volatility $\sigma_o(M)$. Then, $T_{s,o}(\cdot)$ and $T_{h,o}(\cdot)$ should solve the following ODEs:  
    \begin{align}
        -1 & = \mu_o(M) T^{\prime}_{s, o}(M) + \frac{1}{2} \sigma_o^2(M) T^{\prime \prime}_{s, o}(M), \quad \text{with} \quad T^{\prime}_{s, o}(\underline{M}_o) = 0, \ T_{s, o}(\overline{M}_o) = 0, \notag \\
        -1 & = \mu_o(M) T^{\prime}_{h, o}(M) + \frac{1}{2} \sigma_o^2(M) T^{\prime \prime}_{h, o}(M), \quad \text{with} \quad T_{h, o}(\underline{M}_o) = 0, \ T^{\prime}_{h, o}(\overline{M}_o) = 0. \notag
    \end{align}
    The Neumann conditions impose instantaneous reflection at the non-target boundary, while the Dirichlet conditions set the hitting time to zero upon arrival at the target boundary.
\end{Proposition}

The result follows from the Feynman-Kac representation for diffusions with reflecting boundaries \citep[see, e.g.,][]{lions1984stochastic, karatzas2014brownian}. For a detailed argument in a closely related setting, see Section 4.2 of \citet{henriet2016dynamics}.

\begin{figure}[tb]
    \centering
    \includegraphics[width=0.45\linewidth]{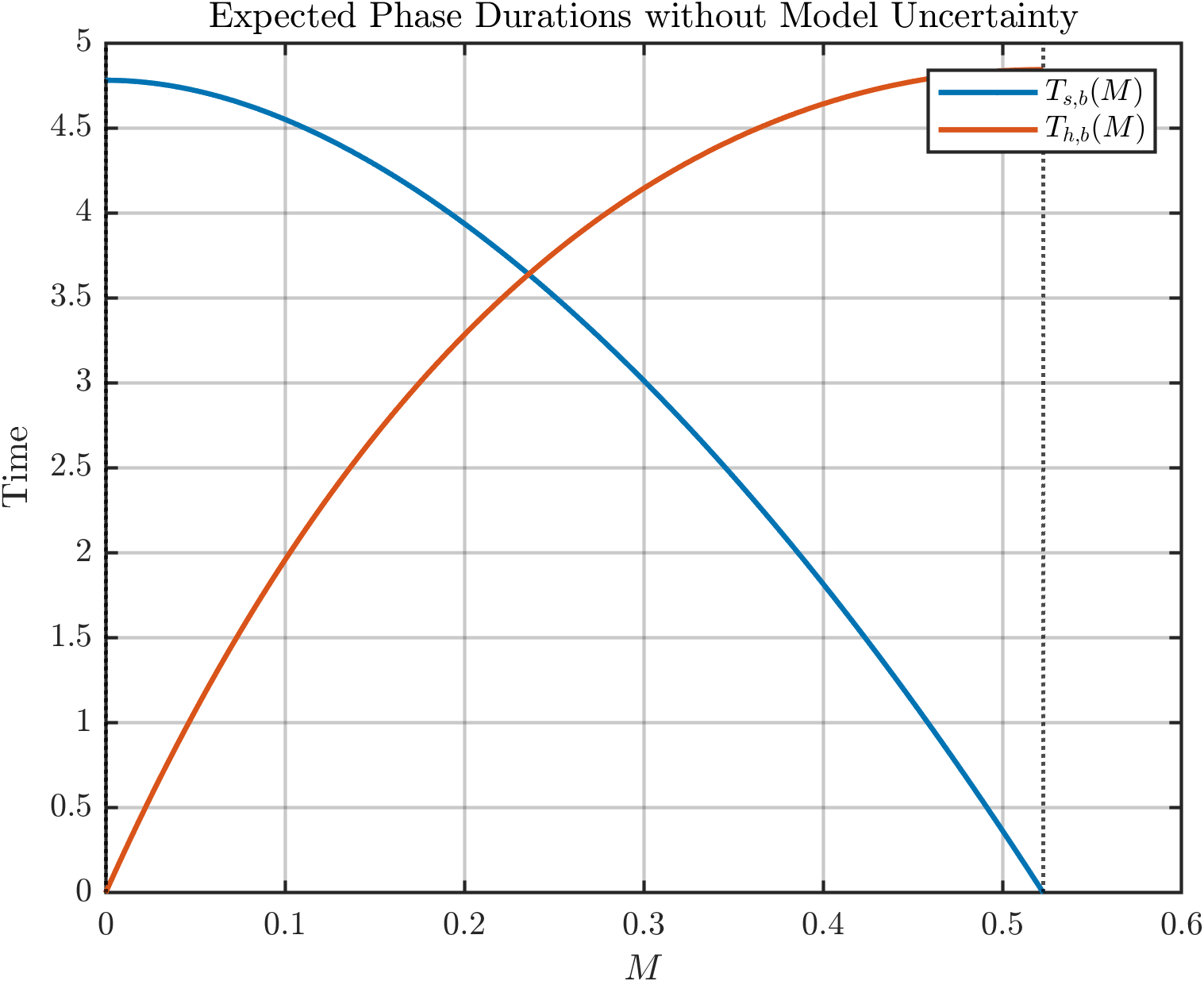}
    \includegraphics[width=0.45\linewidth]{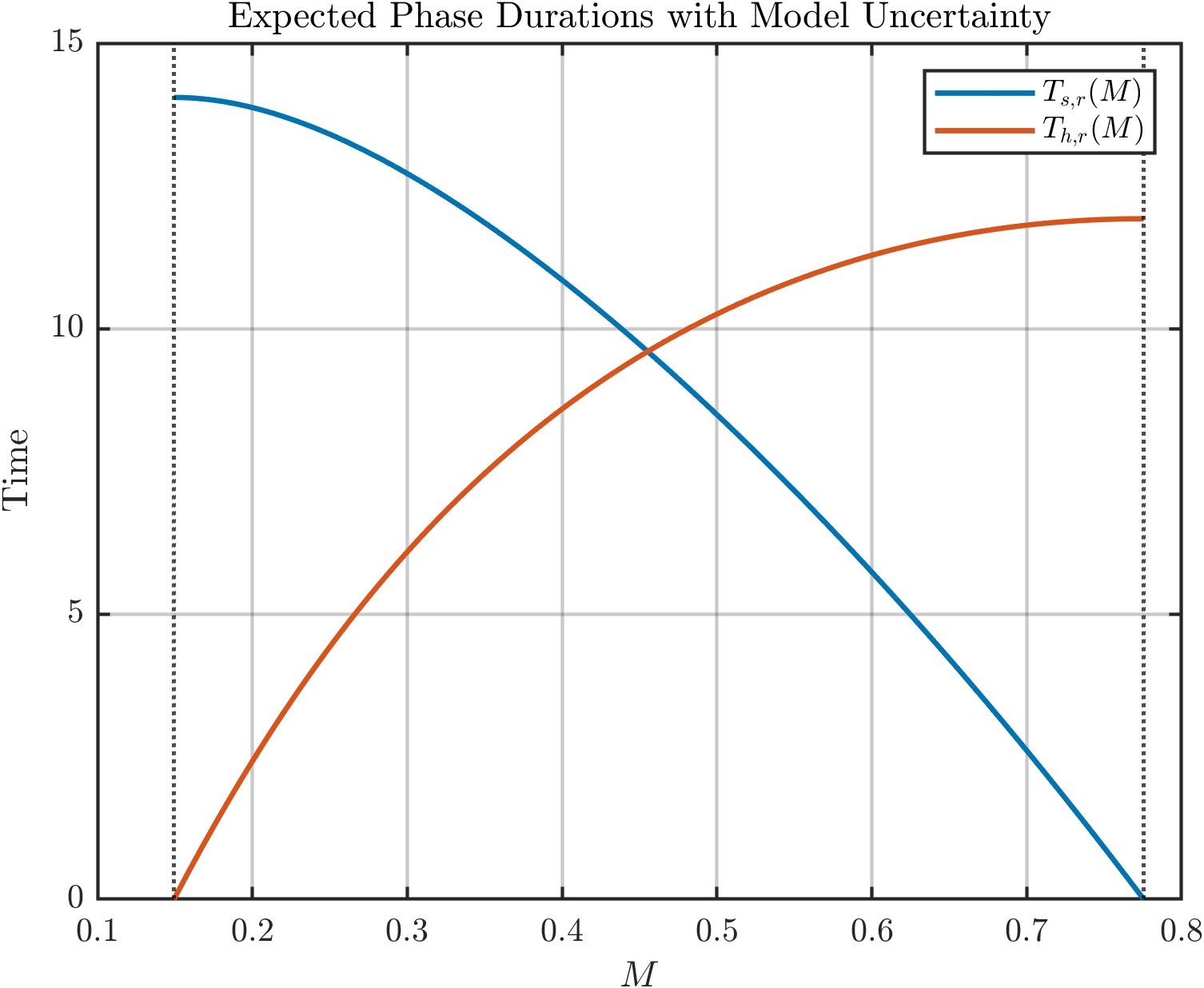}
    \caption{Expected Phase Durations with and without Model Uncertainty.}
    \label{Figure Duration Comparison}
\end{figure}

Figure \ref{Figure Duration Comparison} reports the numerical solutions for $T_{s,o}(\cdot)$ and $T_{h,o}(\cdot)$ under both equilibria. Specifically, we obtain $T_{s,b}(\underline{M}_{b}) = 4.78$, $T_{h,b}(\overline{M}_{b}) = 4.84$, $T_{c,b} = 9.62$, while $T_{s,r}(\underline{M}_{r}) = 14.05$, $T_{h,r}(\overline{M}_{r}) = 11.92$, $T_{c,r} = 25.97$. First, the results indicate that the expected durations of both the soft and hard markets are substantially longer in the robust equilibrium with model uncertainty. As discussed above, this stems from more conservative underwriting and pricing strategies, which lower volatility and widen the admissible capacity range. 

Second, the durations of the two phases become more asymmetric once model uncertainty is introduced. Under plausible parameterizations, the soft market is expected to last significantly longer than the hard market, consistent with the empirical stylized fact highlighted by \citet{henriet2016dynamics}. By contrast, while the equilibrium without model uncertainty in \citet{henriet2016dynamics} also theoretically predicts such asymmetry, this feature is less pronounced in our benchmark results due to differences in the demand specification and parameter choices. 

Table \ref{Table Expected Duration} presents the estimated results of robust market phase durations under different parameters. While Table \ref{Table Boundary Values} shows that the capacity range is non-monotone in the robustness degree $\theta$, here the expected duration of the insurance cycle appears to decrease with $\theta$. The more ambiguity-averse the insurers are, the longer the cycle becomes. This pattern reflects that stronger robustness concerns slow down capital adjustments and dampen fluctuations, thereby prolonging the time it takes for reserves and prices to traverse between boundaries. Similarly, with respect to the external financing cost $\gamma$, higher values of $\gamma$ are associated with longer cycle durations. A higher cost of recapitalization makes insurers more cautious in their capacity management, reducing the speed of adjustment and thereby extending both the soft and hard market phases. 

\begin{Remark}
    In our numerical analysis, all parameters are specified on an annual basis. After incorporating robust pricing and liquidity management strategies, the expected duration of the insurance cycle extends from about 10 years to roughly 26 years, far exceeding the commonly reported 6-10 year range in earlier empirical studies \citep{cummins1987international, harrington2013insurance}. Recent literature has questioned the very existence of underwriting cycles. For instance, \citet{boyer2012underwriting} estimate cycle lengths of 8-11 years using AR(2)/AR(3) models, but these estimates are highly sensitive to small variations in coefficients and, within a 38-year sample, cover only 3-4 cycles, resulting in very low statistical confidence. Our results provide a possible explanation: underwriting cycles may indeed exist, but their duration is much longer than previously assumed. Consequently, short data samples may fail to contain enough realizations to allow for robust statistical inference. 
\end{Remark}

\begin{table}[htbp]
\centering
\setlength{\tabcolsep}{5pt}
\renewcommand{\arraystretch}{1.4}
\small
\begin{tabular}{cccc}
\hline
Parameter       & Soft Market Duration, $T_{s,r}(\underline{M}_{r})$ & Hard Market Duration, $T_{h,r}(\overline{M}_{r})$ & Total Duration, $T_{c,r}$ \\ \hline
$\theta = 0.5$  & 68.23                                              & 37.33                                             & 105.56                    \\
$\theta = 2.8$  & 14.05                                              & 11.92                                             & 25.97                     \\
$\theta = 5$    & 10.98                                              & 9.85                                              & 20.83                     \\
$\theta = 50$   & 6.14                                               & 6.11                                              & 12.25                     \\
$\theta = 500$  & 5.03                                               & 5.10                                              & 10.13                     \\ \hline
$\gamma = 0.06$ & 6.27                                               & 5.30                                              & 11.57                     \\
$\gamma = 0.1$  & 8.87                                               & 7.44                                              & 16.31                     \\
$\gamma = 0.2$  & 14.05                                              & 11.92                                             & 25.97                     \\
$\gamma = 0.3$  & 18.23                                              & 15.83                                             & 34.06                     \\
$\gamma = 0.4$  & 21.83                                              & 19.43                                             & 41.26                     \\ \hline
\end{tabular}
\caption{Numerically Solved Expected Durations.}
\label{Table Expected Duration}
\end{table}

\subsection{Ergodic Property}

For the insurance cycle driven by fluctuations in aggregate capacity, we can further study its ergodic property. In other words, we are interested in whether the capacity process $M_t$ with reflecting boundaries $[\underline{M}, \overline{M}]$ converges to a stationary distribution in the long run, so that insurance cycles can be regarded as statistically recurrent stochastic fluctuations. 

\begin{Proposition}
Let $o\in\{b,r\}$ index the benchmark and robust equilibria, Then $\{M_{o,t}\}_{t\geq 0}$, evolving according to \eqref{Equilibrium M Dynamics Without} or \eqref{Equilibrium M Dynamics With}, is a strong Markov diffusion that admits a unique invariant probability measure $\pi_o$ and is ergodic in the sense that, for every bounded measurable $f$: 
\begin{equation}
    \lim_{T \rightarrow \infty} \frac{1}{T} \int_{0}^{T} f(M_{o, t}) \mathrm{d}t = \int_{\underline{M}_{o}}^{\overline{M}_o} f(M) \pi_o(\mathrm{d}M), \quad \text{a.s.}. \notag
\end{equation}
Moreover, $\pi_o$ has a density $\tilde{\pi}_o(\cdot)$ given by: 
\begin{equation}
    \tilde{\pi}_o(M) = \frac{\kappa_o }{\big(D(p_o^{\ast}(M)) \eta \big)^2 } \exp \left( 2 \int_{\underline{M}_o}^{M} R_o^{\ast}(z) \mathrm{d}z  \right), \quad M \in [\underline{M}_o, \overline{M}_o], \notag
\end{equation}
where the normalizing constant is $\kappa_o^{-1}=\int_{\underline M_o}^{\overline M_o} \frac{1}{\big(D(p^{\ast}_o(M))\eta\big)^2} \exp\!\Bigg( 2\int_{\underline M_o}^{M} R^{\ast}_o(z) \mathrm dz \Bigg) \mathrm dM$. 
\end{Proposition}

The stationary density in the proposition is derived from the Fokker-Planck equation. Importantly, the term $R(M) = \frac{\mu(M)}{\sigma^2(M)}$ appears as the key driver inside the exponential, playing a role analogous to a kernel that shapes the distribution. 

\begin{figure}[t]
    \centering
    \includegraphics[width=0.45\linewidth]{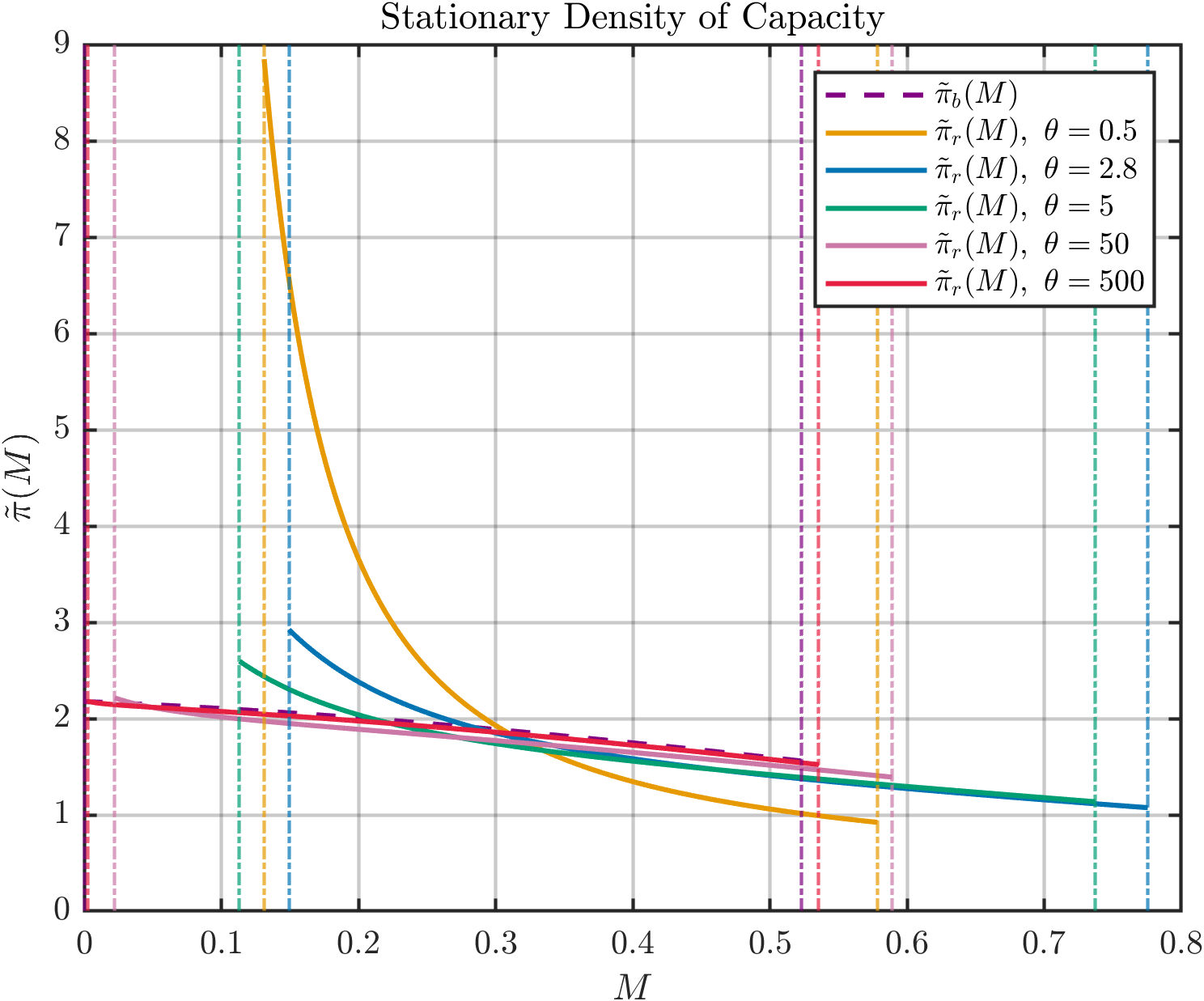}
    \includegraphics[width=0.45\linewidth]{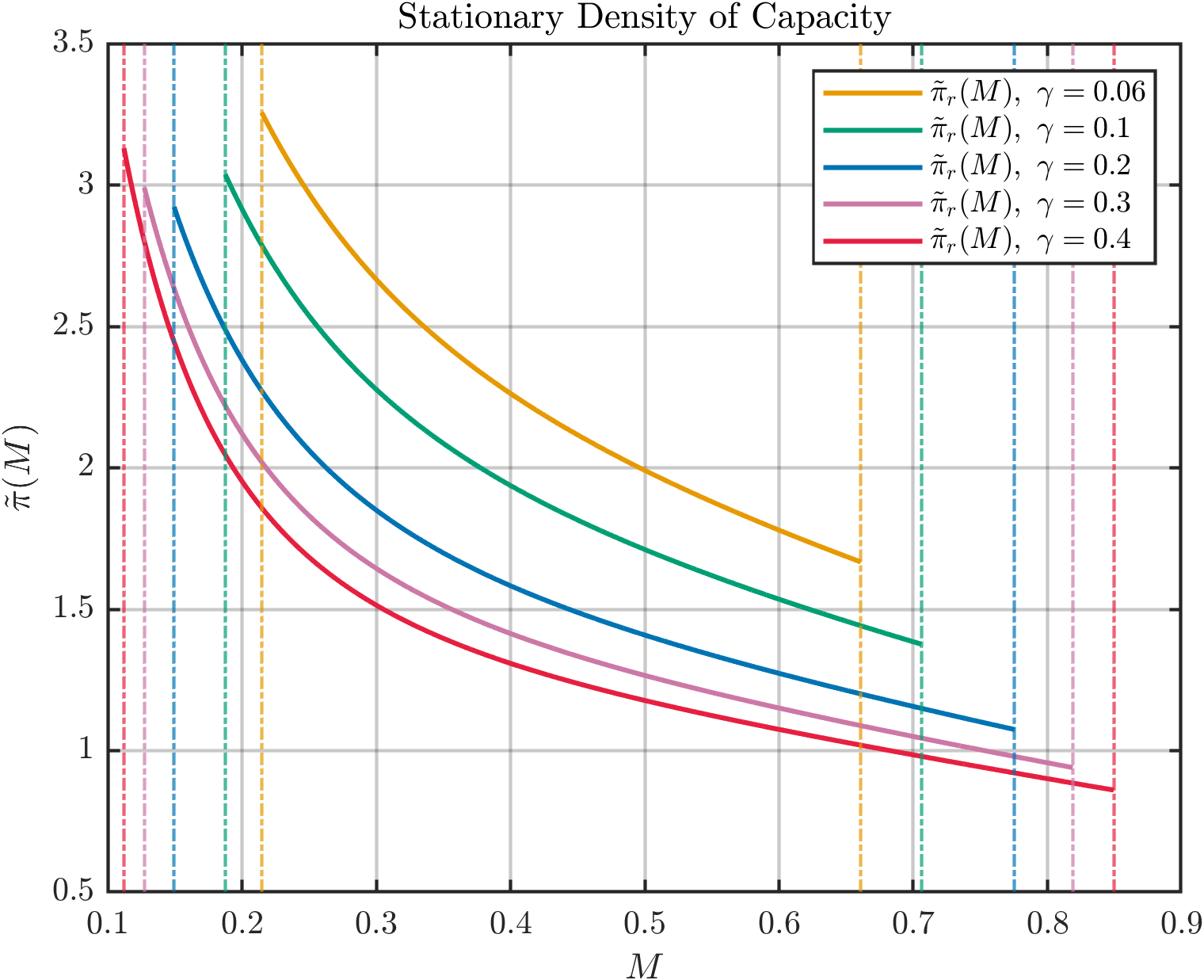}
    \caption{Stationary Density of Capacity with Different Parameters.}
    \label{Figure Stationary Density}
\end{figure}

Figure \ref{Figure Stationary Density} reports the numerically solved results. In both the benchmark and robust equilibria, the density function exhibits a downward slope, indicating that the probability mass is concentrated at relatively low capacity levels. This finding differs from the hump-shaped pattern documented by \citet{luciano2022fluctuations}, which arises when external investment opportunities are considered, but it is consistent with the results of \citet{henriet2016dynamics}.

Moreover, the curves of $\tilde{\pi}_b$ and $\tilde{\pi}_r$ intersect, showing that in the robust equilibrium the density is higher at low capacity levels and lower at high capacity levels. As ambiguity aversion strengthens, the stationary distribution shifts further toward the left boundary, reflecting more conservative capital management in the insurance sector. This also implies that when the industry is hit by a severe negative shock that substantially depletes reserves, it becomes harder and takes considerably longer for insurers to escape from such low-capacity states. Finally, an increase in the external financing cost $\gamma$ raises the marginal value of precautionary reserves, thereby flattening the distribution and spreading the density more evenly across a wider range of capacity levels. 

\begin{Remark}
    \citet{boyer2012underwriting} show that using estimated cyclical components to forecast underwriting ratios 1-2 years ahead yields very poor accuracy. Our results provide a structural explanation: the distribution of insurance cycles is not uniform across states. Once model uncertainty is introduced, the capacity process becomes more skewed toward low-capacity states, where insurers tend to remain trapped for longer and recover more slowly. This persistence in depressed states implies that, even though cycles exist in theory, their short-run predictive content is extremely limited. 
\end{Remark}

\section{Conclusion} \label{Section Conclusion}

This paper investigates how model uncertainty influences equilibrium insurance pricing and liquidity management in a dynamic competitive market. By incorporating insurers' robustness preferences, we show that ambiguity aversion fundamentally reshapes underwriting behavior, leading to higher premiums, high market-to-book ratios and more conservative liquidity management. The equilibrium outcomes reveal several novel patterns: the admissible capacity range expands, the expected length of underwriting cycles is substantially prolonged, and the stationary distribution of reserves becomes more concentrated in low-capacity states. Together, these features suggest that insurers adopt more cautious strategies when facing model misspecification, but at the cost of reduced market activity and longer recovery times following adverse shocks. 

These findings provide theoretical justification for regulatory interventions such as stricter reserve requirements and more robust solvency standards, which can enhance the resilience of the insurance sector. More broadly, our results offer a potential explanation for the mixed empirical evidence on underwriting cycles: cycles may indeed exist, but are much longer than commonly assumed, making them difficult to detect within conventional data samples. 

There are several ways to refine our model to explore more meaningful questions. For instance, we assume that insurers are homogeneous, differing only in their initial capital levels, which leads them to adopt identical underwriting and pricing strategies. However, in reality, insurers exhibit significant heterogeneity, and capital adequacy plays a crucial role in determining firm size and market power. Larger and smaller insurers are likely to adopt different strategies, making the analysis of market structure and its implications a promising avenue for future research. 

Building on this, another important yet overlooked factor in our model is the optimal liquidation decision, which is commonly studied in liquidity management problems. In our current setting, this omission is not a major concern, as insurers are assumed to be homogeneous, and the entire insurance sector does not face insolvency. However, in a setting with heterogeneous insurers, model uncertainty in emerging or catastrophic risks could disproportionately impact smaller insurers, making them more susceptible to insolvency. Investigating how heterogeneous insurers respond to model uncertainty, including potential liquidation decisions, could further enrich the understanding of robust insurance pricing and liquidity management strategies.

\newpage

\bibliographystyle{apalike} 
\bibliography{aguiar} 

\end{document}